%% file: aranzanaresubmission_version.tex
\documentclass[useAMS,usenatbib]{mn2e}

\usepackage{newtxtext,newtxmath}
\usepackage{pdflscape}
\usepackage{supertabular}
\usepackage[T1]{fontenc}
\usepackage{ae,aecompl}

\usepackage{amsmath}	
\usepackage{amssymb}	
\usepackage{gensymb}
\usepackage{textcomp}
\usepackage{graphicx,subfigure,caption}
\usepackage{url}
\usepackage{multicol}
\usepackage{longtable}
\pdfoutput=1

\include{ref_abbrevs}
\title[Optical variability properties of {\it Kepler}/K2 AGN]{Short time-scale optical variability properties of the largest AGN sample observed with {\it Kepler}/K2 }

\author[E. Aranzana et al.]{
E. Aranzana,$^{1}$\thanks{E-mail: E.Aranzana@astro.ru.nl}
E. K\"ording,$^{1}$
P. Uttley,$^{2}$
S. Scaringi$^{3}$
and S. Bloemen$^{1,4}$
\\
$^{1}$Department of Astrophysics/IMAPP, Radboud University, P.O. Box 9010, 6500 GL Nijmegen, The Netherlands\\
$^{2}$Anton Pannekoek Institute for Astronomy, University of Amsterdam, Science Park 904, NL-1098 XH Amsterdam, The Netherlands\\
$^{3}$Department of Physics and Astronomy, University of Canterbury, Private Bag 4800, Christchurch, New Zealand\\
$^4$NOVA Optical InfraRed Instrumentation Group, Oude Hoogeveensedijk 4, 7991 PD Dwingeloo, The Netherlands}

\date{Accepted 2018 February 13. Received 2018 February 5; in original form 2017 May 19}

\pubyear{2018}

\begin{document}
\label{firstpage}
\pagerange{\pageref{firstpage}--\pageref{lastpage}}
\maketitle

\begin{abstract}

We present the first short time-scale ($\sim$hours to days) optical variability study of a large sample of Active Galactic Nuclei (AGN) observed with the {\it Kepler}/K2 mission. The sample contains 252 AGN observed over four campaigns with $\sim 30$ minute cadence selected from the Million Quasar Catalogue with R magnitude $< 19$. We performed time series analysis to determine their variability properties by means of the power spectral densities (PSDs) and applied Monte Carlo techniques to find the best model parameters that fit the observed power spectra. A power-law model is sufficient to describe all the PSDs of our sample. A variety of power-law slopes were found indicating that there is not a universal slope for all AGN. We find that the rest-frame amplitude variability in the frequency range of $6\times10^{-6}-10^{-4}$ Hz varies from $1-10$ \% with an average of 1.7 \%. We explore correlations between the variability amplitude and key parameters of the AGN, finding a significant correlation of rest-frame short-term variability amplitude with redshift. We attribute this effect to the known ``bluer when brighter'' variability of quasars combined with the fixed bandpass of Kepler data. This study also enables us to distinguish between Seyferts and Blazars and confirm AGN candidates. For our study we have compared results obtained from light curves extracted using different aperture sizes and with and without de-trending. We find that limited de-trending of the optimal photometric precision light curve is the best approach, although some systematic effects still remain present.
\end{abstract}




\section{Introduction}

Active galactic nuclei (AGN) are powered by accretion onto supermassive black holes in the centres of the galaxies. The  material forms a geometrically thin, optically thick accretion disc that emits mostly in the optical/ultraviolet regime. Considerable efforts have been directed at studying AGN, trying to constrain the physical parameters of these sources and to understand the accretion disc physics. Since AGN are extremely distant and compact, direct optical imaging is not possible and indirect methods are required to investigate their behaviour. Reverberation mapping has proved to be a useful tool to estimate the black hole mass, using the Doppler broadening of the emission lines and the distance to the broad line region \citep[e.g.][]{ReverberationPeterson}. The latter is measured from the light travel time delay of the continuum emission to the broad line region. Another common indirect method applied to these sources is the time-series analysis of their light curves, since AGN present intrinsic variability observed at different wavelengths from radio to X-ray and gamma-ray and on a broad range of time-scales from hours to years \citep[e.g.][]{Mushotzky2011}. Using timing techniques we can give insight into the characteristic time-scales and potentially constrain disc parameters such as the viscosity parameter, $\alpha$ \citep[][]{Starling2004}. 

\begin{table*}
\begin{minipage}{\textwidth}
 \centering
 \caption{Observations of AGN with {\it Kepler}/K2 mission at $30$ min cadence during four campaigns. It indicates the coordinates of the center of the field of view, the duration of the observation, the date and the number of AGN observed in each field.}
 \label{table:obs}	
	 \begin{tabular}{lccccc}
		\hline\hline
		Field & RA & Dec & Start & Stop & Sources\\
		\hline
		0 & 06:33:11 & +21:35:16 & 2014/03/12 & 2014/05/27 & 37 \\ 
		1 & 11:35:46 & +01:25:02 & 2014/05/30 & 2014/85/21 & 149 \\ 
		2 & 16:24:30 & -22:26:50 & 2014/08/23 & 2014/11/13 & 13 \\ 
		3 & 22:26:40 & -11:05:48 & 2014/11/14 & 2015/02/03 & 76 \\ 
		\hline
		\end{tabular}
\end{minipage}

\end{table*}
 
Long-term optical variability studies have been performed in the past decades, mainly dedicated to determine the properties of a well defined sample, i.e. the Palomar-Green quasars observed with the WISE observatory \citep[][]{Giveon1999Palomar} or the Sloan Digital Sky Survey (SDSS) \citep{vandenBerk2004,Sesar2007SDSS,MacLeod2010}. The ground-based optical observations have the disadvantage of being subject to the atmospheric turbulence, which can introduce photometric errors (see comparison of {\it Kepler} and ground-based telescopes in \citet{Mushotzky2011}) hiding the intrinsic variability from the source, especially on short time-scales where variability amplitudes are small. These optical variability studies have been also limited by the sparse sampling, with only a few hours of continuous short cadence observations.
Recently, there have been a small number of studies on short-term variability using {\it Kepler} observations, mostly focused on bright sources situated in our local Universe. This mission can provide unprecedented accurate and continuous short cadence light curves, excellent for time-series analysis. \citet{Mushotzky2011} studied 4 AGN observed with {\it Kepler} and determined that the power spectral densities (PSDs) are consistent with a power-law fit with slopes of $-2.6$ to $-3.3$, higher than those seen in X-ray PSDs. More recently, \citet{Edelson2014} studied the AGN Zw $229-15$ discovering a 5 day break in the power spectrum associated to a characteristic time-scale of the system. The latter can be associated with either the dynamical or the thermal time-scale for an $\alpha\sim0.1$ and an emission distance of $100-1000$ Schwarzschild radii. \\

Given the limited studies on optical short-term variability of AGN, well sampled observations and an unbiased sample of sources are needed to give insight into the physical processes generating the optical fluctuations. Moreover, there have been many studies focused on blazars, where the optical variability is thought to originate in the shocks produced by the relativistic jet, and not many on radio quiet quasars where the optical variable emission comes from the accretion disc \citep[][]{Cellone2007blazars,Carini2011blazar,Ruan2012blazars,Edelson2013blazar}. Therefore, a larger sample of AGN containing objects of different classes with different luminosities and at different distances is required to probe the short-term variability of the population of AGN in the Universe. \

In this work we present a flux-limited catalogue of 252 AGN observed with the {\it Kepler}/K2 mission containing various classes of AGN, the majority situated in our local Universe up to redshift $\sim$4. We explore their power spectral densities to determine their variability amplitudes and their true shape. This is the first large statistical study of AGN on optical short-term variability with a sample containing hundreds of sources. The source selection, the observations and the data extraction are given in Section 2; the time-series analysis methods and simulations are described in Section 3; and the results and the discussion are reported in Section 4 and 5, respectively. \\

\section{Observations and sample selection}

\subsection{The K2 sample of AGN}
We performed observations of a selected sample of AGN with the K2 mission \citep{Howell2014}, the new re-devised mission of {\it Kepler}. With its large field of view $\sim\,105\,\rm deg^{2}$, {\it Kepler} can monitor hundred of thousand of sources every 29.4 minutes with a duty cycle $>90\,\%$. The {\it Kepler} photometer utilizes one broad bandpass, ranging from 420 to 900 nm. In the current K2 mission each campaign consists of monitoring a unique region of the sky for $\sim$80 days, performing both long cadence observations of 29.4 minutes as well as short cadence $\sim$1 minute observations of a limited number of targets. The K2 mission observes thousands of sources in each field in long cadence and tens of sources in short cadence mode, less than the original mission. \\ 
\begin{figure*}
\includegraphics[width=7.5cm]{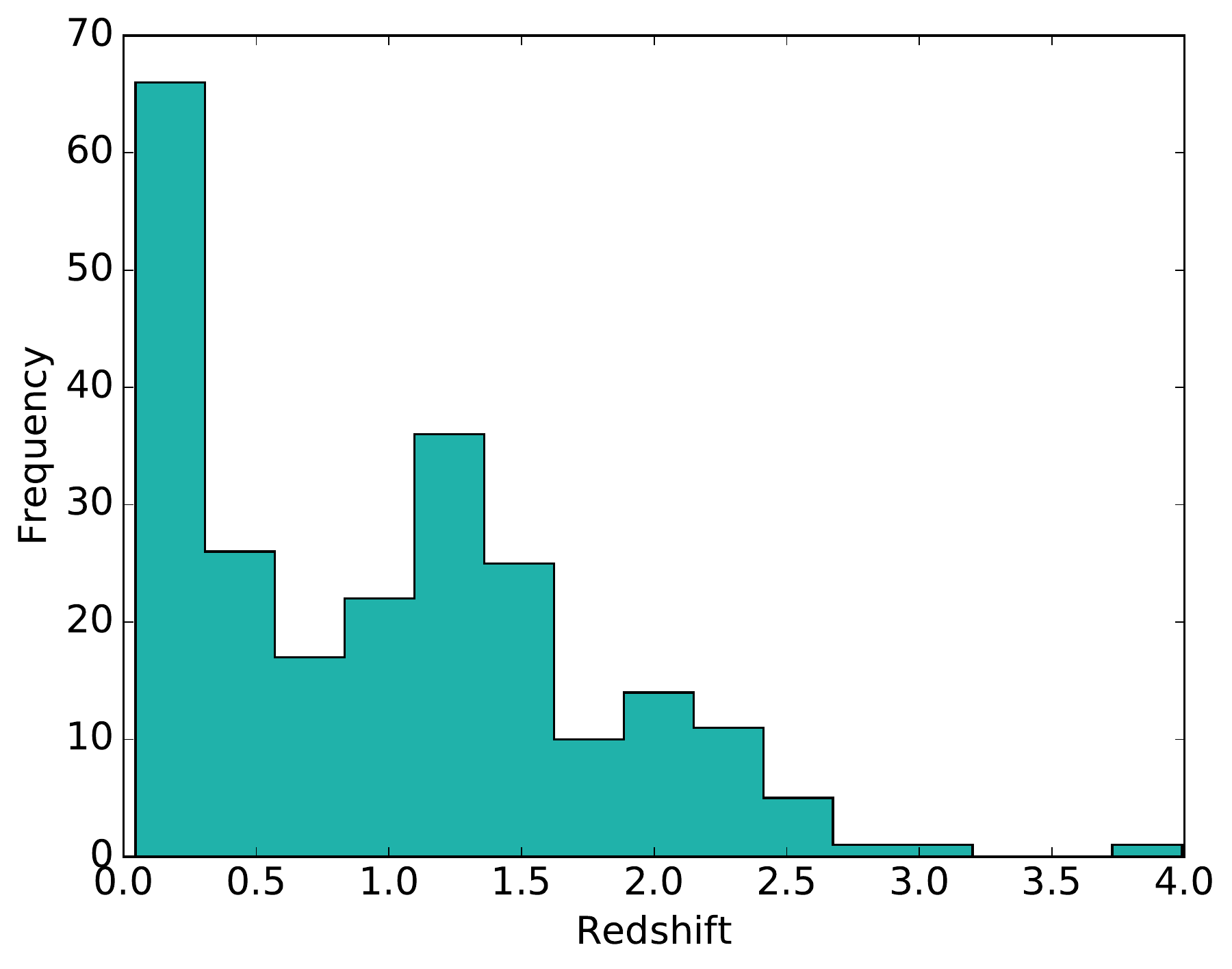}
\includegraphics[width=7.5cm]{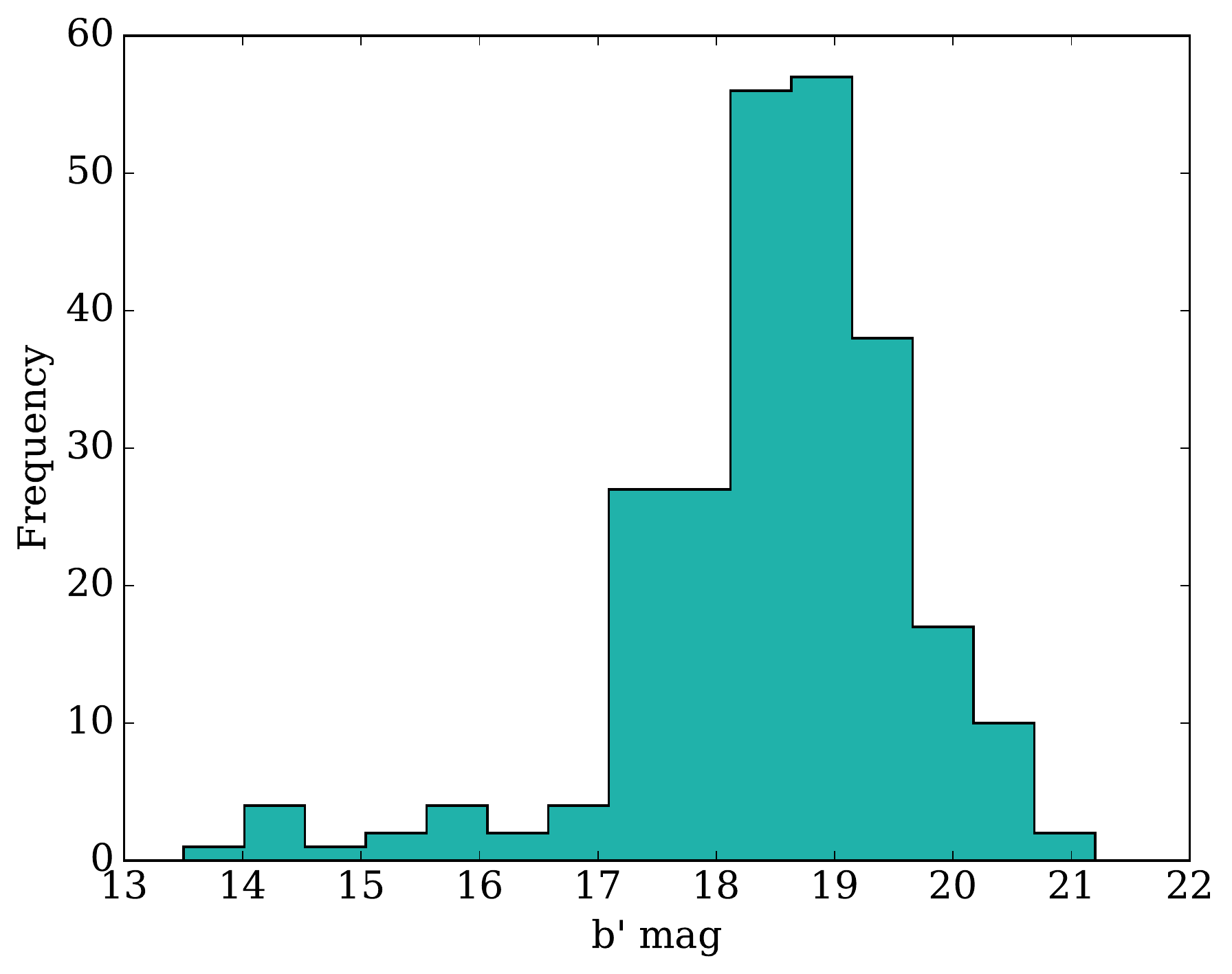}
\centering
\caption{{\it Left:} Redshift distribution of the sample, where a large number of sources are in our local Universe. {\it Right:} Distribution of the $b'$ magnitudes, the majority have a brightness of $\sim19$ mag.}
\label{Fig: histoz}
\end{figure*}
\begin{figure}
\includegraphics[width=6cm]{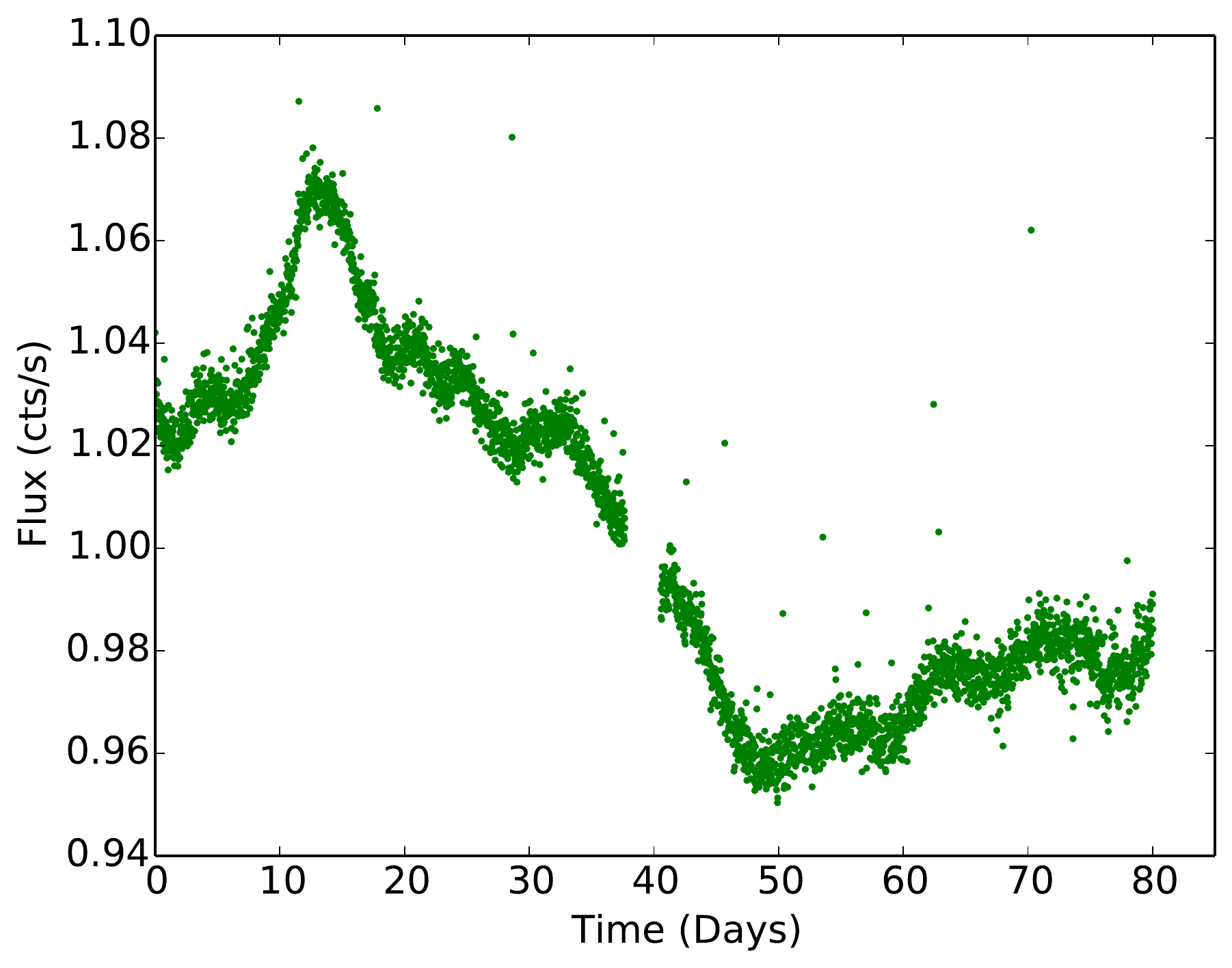}
\includegraphics[width=6cm]{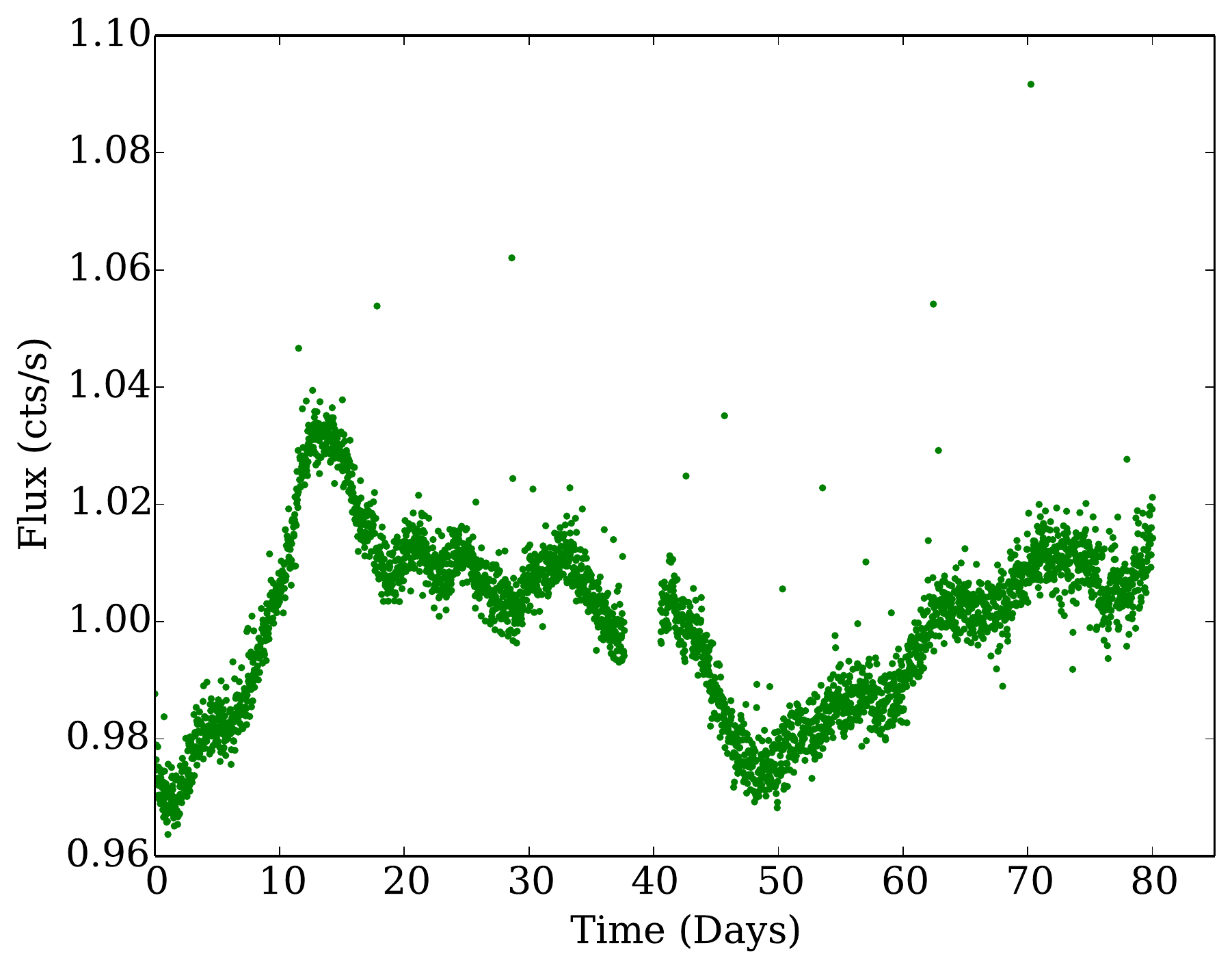}
\includegraphics[width=6cm]{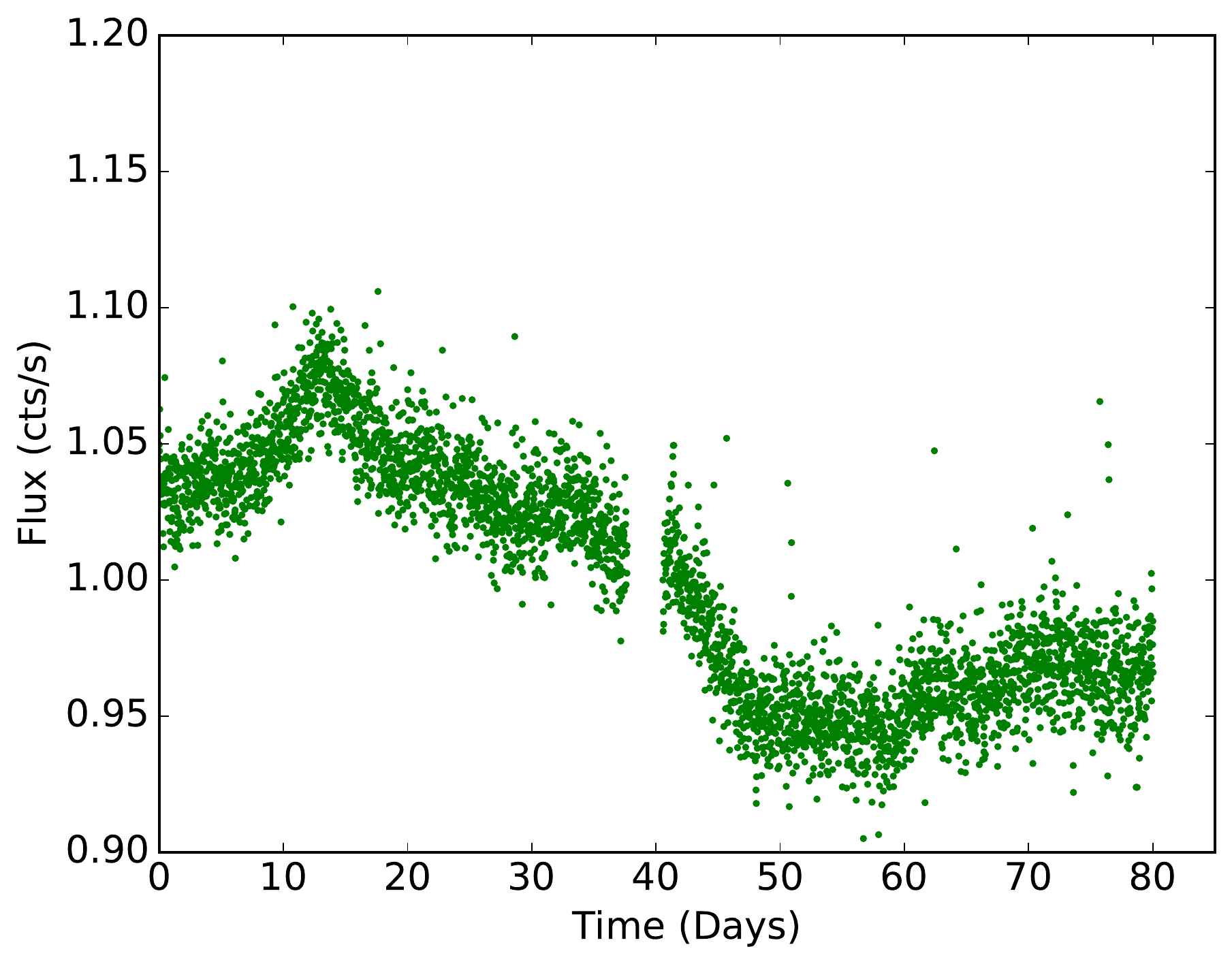}
\centering
\caption{{\it Upper:}{\it Kepler}/K2 `optimal' light curve of EPIC number 201805756 (SDSS $J112328.12+052823.2$). The light curve spans over 80 days with a gap when the telescope turned to transmit data. {\it Middle:} Same light curve de-trended. {\it Lower:} Largest aperture extracted light curve using 66 pixels where the S/N ratio is much lower than the optimal light curve (8 pixels extraction).}
\label{Fig: lightcurve1}
\end{figure}

By selecting only sources with a known R band magnitude brighter than 19, the probability of being a quasar is greater than 75\%, yielding 547 candidates. We proposed to observe the 200 brightest Quasars that fall on silicon (K2fov flag = 2) from the ``Million Quasars Catalogue, Version 3.7$"$ \citep[see][]{Flesch2013}. The first campaign (C0) was implemented to prove that the mission was still viable, but it was not pointing accurately at the beginning of the campaign. This caused the available data to only cover $\sim$35 days. Only 49 AGN candidates were observed, but more observations were needed to statistically link the variability properties with parameters such as the black hole mass, the redshift and the luminosity. For that reason we pursued monitoring during campaigns C1, C2 and C3 to observe more quasars. The final sample consists of 252 AGN observed during campaigns 0 to 3, including radio-galaxies, Blazars, Seyfert 1 and Seyfert 2 (see Table \ref{table:obs}). The majority of the sources were selected from the SDSS, thus most of them have known redshifts and masses. The distribution of redshifts and the b$'$ magnitudes of our sample are shown in Fig. \ref{Fig: histoz}. For the majority of the sources used here the b$'$ magnitudes listed in the Million Quasar catalogue were extracted from POSS-I \citep{POSS1}. \\

\subsection{Light curve extraction and systematic effects}\label{Sec:lc}

 We used the Mikulski Archive for Space Telescopes portal (MAST) to retrieve the K2SFF light curves \citep{Vandenburg2014}. These light curves have been corrected for the photometric variability caused by jitter in the precise pointing of the spacecraft using a self-flat-fielding (SFF) approach. This correction improves the photometric precision by a factor $2-5$. In addition, data points taken when Kepler fires its thrusters are removed \citep[see][for details]{Vandenburg2014}. 

Two types of aperture are considered: a circular aperture around the target and a region defined from the telescope's pixel response function (PRF) \citep{Bryson2010PRF}. For both types of aperture, light curves were extracted for a series of different aperture sizes and the photometric precision of each light curve was measured using the Combined Differential Photometric Precision (CDPP) metric, similar to the one used by the {\it Kepler} pipeline. This metric measures the photometric stability over a 6-h period. The aperture that provides the best photometric precision in this metric is selected as the `optimal' aperture. This `optimal' aperture is designed to be ideal for planet searches, but is not necessarily ideal for other variability studies. 

The K2 light curves suffer from a series of systematic effects that could affect our study on short-time variability of AGN, and the degree to which light curves are affected is strongly dependent on the aperture size. Apart from the pointing jitter mentioned above, which can mostly be corrected for, there are two important systematics that are difficult to account for and that can strongly affect the light curves on longer time-scales: the {\it differential velocity aberration} (DVA) and the Moir\'e effect. The latter is caused by the crosstalk between the four fine guidance sensors channels and the 84 readout channels. Also a high-frequency oscillation that arises from an instability of the amplifiers generates a time-varying Moir\'e pattern and a rolling band, seen in the light curves as ripples with time-scales of days. The channels affected by the rolling band are listed in Table 13 of Kepler Archive Manual \footnote{https://archive.stsci.edu/kepler/manuals/KSCI-19033-001.pdf}. Since there is no procedure to mitigate this effect we will further indicate the light curves of this study that can be affected by the rolling band \citep{Clarke2014}.

The DVA originates from the annual motion of the spacecraft around the sun. While observing a certain region of the sky, the angle with respect to the velocity vector of the spacecraft changes, which causes the target point spread function (PSF) to move across the detector at the sub-pixel level. Since the aperture that is used never completely captures all the flux in the PSF, this drift causes the fraction of flux captured by the selected pixels to change with time. This is translated into falling or rising slopes in the light curve, showing a variability in the quarter that is not intrinsic to the source.

The DVA effect on {\it Kepler} light curves has been pointed out in previous studies of AGN, where different authors used different approaches to alleviate this effect. An option commonly used is to correct the light curves using co-trending basis vectors (CBVs), which represent the most common trends found in each channel. The main problem of using them, is that there is not a clear way to ensure that we are only removing the artefacts and not the intrinsic variability of the system in study \citep[see][]{KinemuchiDVA}. Another approach has been to remove the linear trend from the whole light curve by ``end-matching'' \citep{Mushotzky2011,Wehrle2013}. In general, de-trending the light curves has to be done with caution because of the inherent risk of removing intrinsic variability. 

The number of pixels that were stored per target on board of the spacecraft is much larger during K2 than during the nominal {\it Kepler} mission. Larger apertures can therefore be used. As the fraction of flux near the edges of the aperture is much smaller for larger apertures, which extend further into the wings of the PSF, light curves extracted using larger apertures are less affected by the DVA effect. The drawback of using larger apertures is an increase in noise caused by additional background light and readout noise from the CCD, and a higher chance that the light curve gets contaminated by light from other nearby sources (so-called `third light').

Therefore, using relatively small apertures is advantageous if it is possible to remove most of the instrumental trends in the light curves without significantly influencing the results of the analysis. For the analysis in the this paper, we used three different light curves for each source:
\begin{itemize}
\item the `optimal' light curve: the light curve selected as optimal in \citet{Vandenburg2014}.
\item a de-trended light curve: the `optimal' light curve de-trended by dividing the light curve by a sine curve with a fixed orbital period of 372.53\,d (the period of motion of the satellite around the sun) and a fitted phase and amplitude.
\item PRF9: the light curve with the largest PRF-shaped aperture available in \citet{Vandenburg2014}. This light curve is the least affected by DVA.
\end{itemize}

An example of the three types of light curves is shown in Fig.~\ref{Fig: lightcurve1}. In Sec.~\ref{Sec:effectDVA} and Appendix~\ref{Ap:trendeffect} we explore the influence of de-trending on the results using simulated light curves. We find that trends significantly influence the derived spectral slope and that de-trending allows one to recover the original slope. We also show that for a large sample, the differences between large apertures and de-trended `optimal' aperture light curves are not significant.

\begin{figure}
\includegraphics[width=8.5cm]{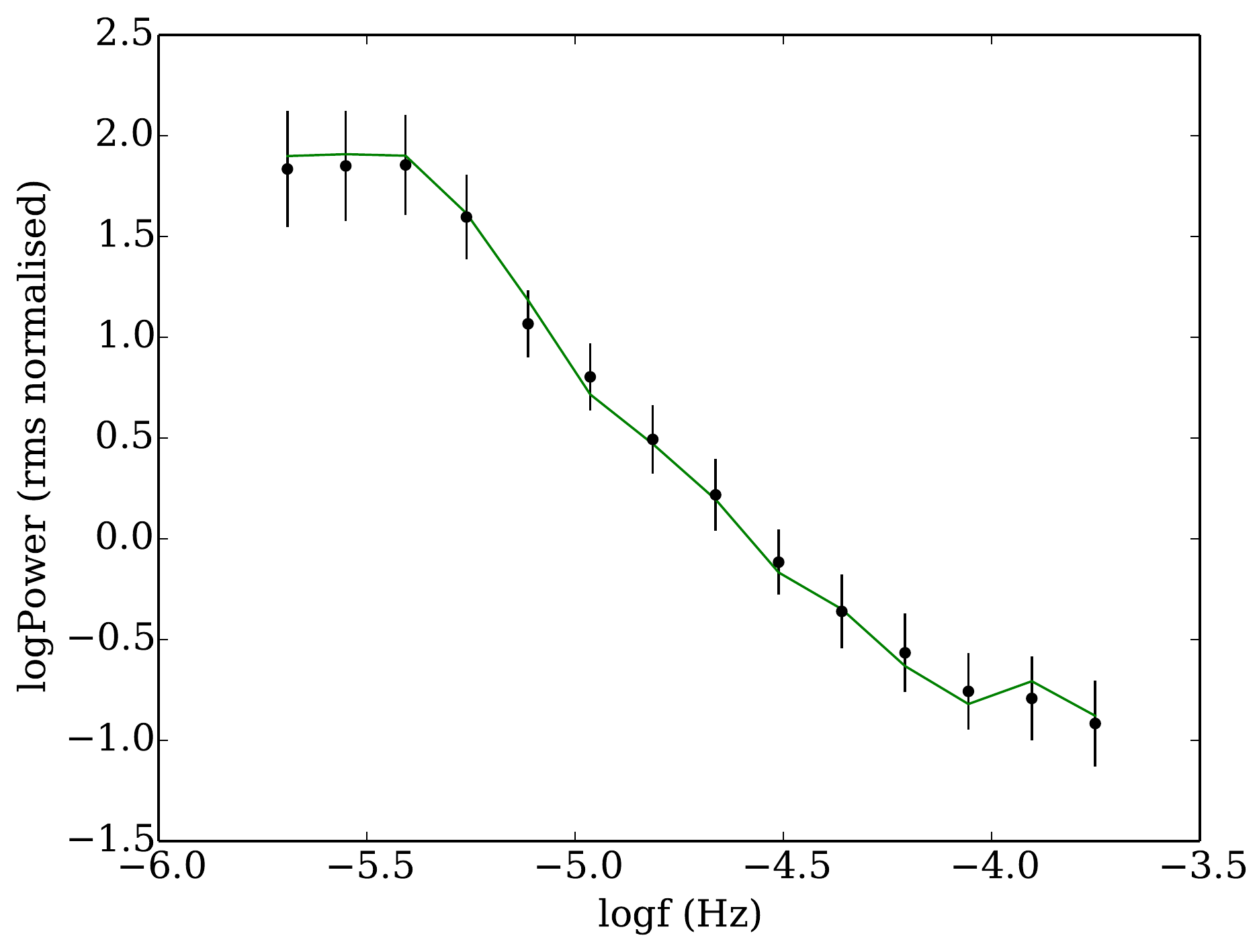}
\centering
\caption{Power spectral density of the quasar 201805756. The green solid line indicates the observed power spectra and the black filled circles the best simulated power spectra with errors. The power-law model fits well the observed power spectrum with a power-law index of $-2.4$ and a confidence of $83\,\%$.}
\label{Fig:psdexample}
\end{figure}

\subsection{Outlier removal}
For each light curve obtained from MAST we have filtered the outliers that can be associated with cosmic rays or systematics from the detector. For that, we calculated the difference between each data point and the next, filtering out data points that were deviating more than two times the standard deviation of the entire light curve. This filtered less than $<1\%$, since we do not want to subtract intrinsic variability from the source. The standard deviation is significantly larger than the expected Poisson noise, as the power spectrum of AGN is dominated by red noise.

\section{Methods}
\label{sec:methods}

The optical AGN light curves show aperiodic fluctuations, for that reason time-series analysis was performed to characterize the broad-band red noise of each system. This technique, that has been extensively applied to X-ray light curves consists in splitting the light curve into segments of equal duration and then computing the power spectra for each of the segments independently. Then, they are averaged and log-binned to obtain the PSDs \citep[see for a full description][]{Papadakis1993}. The {\it Kepler} AGN light curves contain gaps that in some cases may be long, as seen in campaign 1 and 2, where there is a three day gap as {\it Kepler} turned to point its antenna towards the Earth to transmit data (see Fig. \ref{Fig: lightcurve1}). Since the gap is in the middle of the observation we split the light curve in two halves and then each half in 3 segments. In campaign 3 there is no long gap so the light curves are directly split in 6 segments. Thus, for the analysis of these light curves we calculated Lomb-Scargle periodograms \citep{Lomb1976,Scargle1982} on six non-overlapping segments of $\approx$13 days duration. For campaign 0 we only used 3 segments as the duration was shorter than for the other campaigns. Once the observed PSDs for the six segments were obtained we applied the rms normalisation \citep{Miyamoto1991}, and we averaged them. We averaged the logarithm of power in each frequency bin. By doing this the errors will be symmetric and the minimisation function to compare the observed and the modelled power spectra will be linear, so that it will easily converge and find the right set of parameters, thus we obtained $<\log_{10}P_{\rm obs}>$. We measured the white noise that dominates at high frequencies ($> 10^{-4}$ \rm Hz) in the power spectrum to determine the intrinsic PSD. The latter is needed to compute the fractional root-mean square variability (RMS) from each target.  \\

\begin{figure}
\includegraphics[width=8.5cm]{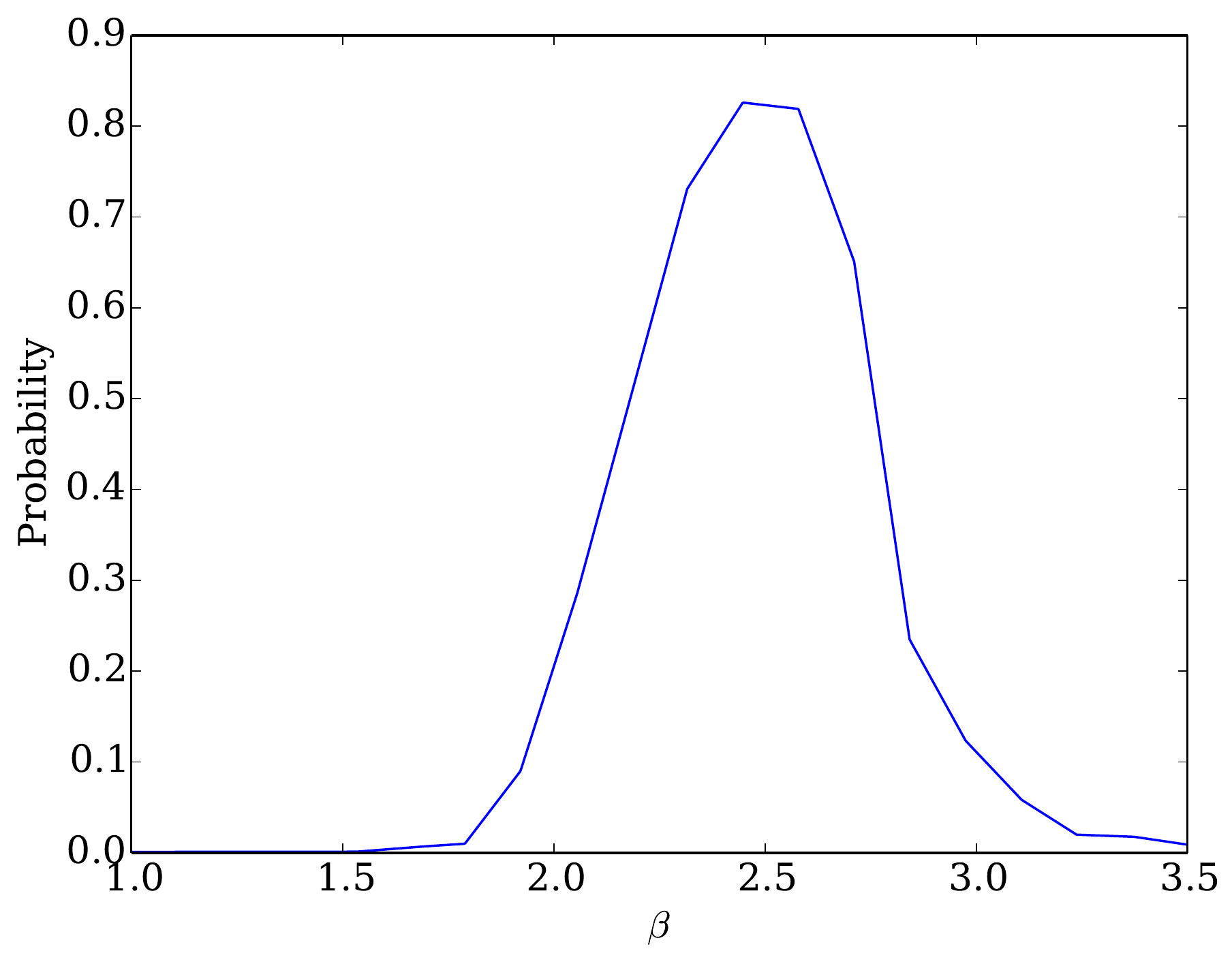}
\centering
\caption{Probability distribution of the different power-law models with power-law indices ranging from $-1.0$ to $-3.5$ for the quasar 201805756, indicating that the slope with the highest probability is $-2.4$.}
\label{Fig:prob}
\end{figure}

Generally, the power spectral densities of AGN present a power-law index of $-2$ at higher frequencies and if the observation lasts long enough a break at lower frequencies can also be observed \citep[e.g.][]{Uttley2002}. The light curves studied in this work have a duration of 80 days, and they present a very strong red-noise-leakage effect, so that the power from low frequencies can leak to high frequencies, and as a result the power spectra look flatter than the true PSD \citep[e.g.][]{vanderKlis1997leakage}. Since this effect changes the real shape of the PSD, in order to fit them we applied a Monte Carlo routine known as {\it PSRESP} developed by \citet{Uttley2002}. The main idea of this method is that by simulating a light curve 1000 times longer than the original one with a custom power-spectral shape and applying the same sampling pattern as the observed light curve we can determine the real power spectral shape.\\

Using the information that we can extract from the observed light curve and its PSD, namely the sampling pattern, the standard deviation and the mean flux, we simulated the light curves using the \citet{Timmerkoenig} routine. This algorithm takes a custom power spectrum model, i.e. broken power-law, power-law, and  derives a light curve with the corresponding power spectrum shape provided as input by the user. {\it Kepler} AGN light curves are not ideal to detect the frequency breaks detected in previous works with ground based telescopes as these observations have a duration of 80 days \citep{Mushotzky2011}. Thus, we created the simulated light curves using a simple power-law model with a power-law index $\beta$. We did notice a bend at low frequencies but this is an artificial effect caused by the Lomb-Scargle routine (see Fig. \ref{Fig:psdexample}). As a test we performed a Fast-Fourier Transform on the same data and the flattening was not observed. After choosing the model with a certain $\beta$ we generated a light curve 1000 times longer than the observed one. In order to save computing time we extended the length of the light curve to the next greater power of 2, since the Fast-Fourier Transform is more efficient. Once the long light curve was generated we chose only a section of the light curve that is 1000 times longer than the original one and we discarded the rest. Then we split the resulting light curve into 1000 segments that have exactly the same length as the original light curve. After applying the observed sampling pattern to each one of those, we computed the 1000 PSDs treating their light curves exactly in the same way as the observed PSD (splitting in segments, applying Lomb-Scargle, normalising it, averaging and log-binning).

We obtained the averaged simulated PSD by averaging the 1000 PSDs, $<\log_{10}\rm P_{\rm mod}>$, and then we minimised the distorted model $\chi^{2}$ statistic, $\chi^{2}_{\rm dist}$, to find the best normalisation and the noise for the power spectrum. The simulated PSD error is computed as the root-mean-square spread about the mean of the simulated 1000 PSDs that have been logarithmically binned, $\delta\rm P_{\rm mod}$.

\begin{equation}
\chi^{2}_{\rm dist} = \sum_{i=1}^{\rm N}\frac{[\log_{10}(10^{(k+<\log_{10}\rm P_{\rm mod}>)}+\rm C_{\rm noise})-<\log_{10}\rm P_{\rm obs}>]^{2}}{\delta\rm P_{\rm mod}^{2}}
\end{equation}

\noindent Here N is the number of simulated light curves,  $<\log_{10}\rm P_{\rm obs}>$ is the observed PSD, {\it k} is the normalisation and $C_{\rm noise}$ is the constant to account for the Poisson noise. For an optimal performance of the fitting, we decided to remove the first three bins of the PSD at low frequencies, since  each bin contained less than four data points. Having obtained the best normalisation, the Poisson noise level and the minimum $\chi^{2}_{\rm dist}$ we estimated the goodness of the fit comparing with the observed PSD as described in \citet{Uttley2002}. Thus, we obtained the p-values that indicate whether the model describes the data well. We performed this procedure for different power-law models, different power-law indices in a range from $-1.0$ to $-3.5$ in steps of 0.1 to find the model that best fits the observed PSD. The power-law index $\beta$ with the highest acceptance probability will be the best match for the observed power spectrum. The upper and the lower limit for $\beta$ are the ones for which the probability is at 10 \%. Then, the final $\rm P_{\rm sim}$ that best describes the observed power spectrum is computed using the $<\log_{10}\rm P_{\rm mod}>$ with the highest probability and the normalisation k and noise level obtained in the minimisation: \\

\begin{equation}
\log_{10}\rm P_{\rm sim} = \log_{10}(10^{(k+<\log_{10}\rm P_{\rm mod,opt}>)}+\rm C_{\rm noise}
\end{equation}
In addition, the fractional rms from the observed and the simulated PSD was obtained by integrating the intrinsic power spectrum (white noise subtracted) in the frequency range $6\times10^{-6}-1\times10^{-4}$ Hz:  

\begin{equation}
\sigma (\%) = \sqrt{\int_{\nu_{1}}^{\nu_{2}}\rm P_{\rm sim}\rm d{\nu}}
\label{eq: frac}
\end{equation}

\noindent In order to calculate the rms in the quasar rest-frame and thus correct from time dilation we used the approach of \citet{Almaini2000timedilation}.  We used the redshift values extracted from the Million Quasar Catalogue, and for the cases with no estimated value in the literature we used the median of the whole sample that is $0.918$:
\begin{equation}
\sigma_{\rm rest}\,(\%) = (1+z)^{(\beta-1)/2}\, \sigma (\%)
\label{eq: dilation}
\end{equation}

\noindent The fractional rms error is then calculated by error propagation and using the errors derived in the analysis for $\beta$.

\subsection{Effect of the DVA}\label{Sec:effectDVA}
We have explored the effect of an instrumental trend on 1000 artificial independent light curves generated using the method described by \citet{Timmerkoenig}. We chose a a power-law  spectrum model with a slope of $-2$, similar to what have been observed in other AGN. We treated them as real light curves by putting the same sampling pattern as observed in one of the K2 light curves. Then, we applied the PSRESP method to find the best slope $\beta$ that matches the ``observed'' power spectrum. We found that the average slope of the 1000 light curves was indeed $-2$. Then, we introduced a trend in all the light curves similar to the trends observed, and we ran the PSRESP method again. We found that the average $\beta$ is steeper than the original ones, at $\approx-3$. An example of an artificial light curve and the same with a trend are shown in Fig. \ref{Fig:fakelcr}. When de-trending these light curves fitting a sine curve, as we did with the real light curves above, and running again our pipeline we retrieved the average slope of $-2$ (see histogram in Fig. \ref{Fig:simulcom}). More detailed information can be found in the Appendix \ref{Ap:trendeffect}. This shows that de-trended light curves are very similar to the real light curves and that removing a sine curve at only one frequency, in this case the orbital period of the satellite, eliminates the dominant long-term instrumental trend. However, there is still evidence of small spurious signals, e.g. the sudden dip in the light curve at day $\sim65$, as seen in sources 201167738, 201185828, 201189418 and 201207010. These sources were observed with neighbouring CCDs on the edge of the field-of-view, and this area was affected by a third light source (e.g.~ghosting). The parameters derived in the analysis of these AGN are consistent with the rest of the sample.  \\

As explained in Sec.~\ref{Sec:lc} we used three different data sets: `optimal' light curves, de-trended `optimal' light curves and aperture PRF9 light curves. We ran the PSRESP pipeline on all three datasets. The power-law indices obtained using the de-trended `optimal' light curves and the PRF9 light curves are within $1\sigma$ of each other, as shown in Fig. \ref{Fig:significance}. Moreover, the average slope is very similar: $2.2\pm0.5$ and $2.1\pm0.6$ respectively. The PRF9 light curves suffer the least from the DVA effect as explained above. Since the results based on the de-trended `optimal' light curves are very close to those of PRF9, we use de-trended `optimal' light curves for further analysis as they are less affected by Poisson noise. We would like to remark that even though de-trending with a single sine curve was shown to be a safe and very effective option in this case, caution is needed when de-trending because it is possible that intrinsic variability of the system gets removed.

\section{Results}
\begin{figure}
\includegraphics[width=8.5cm]{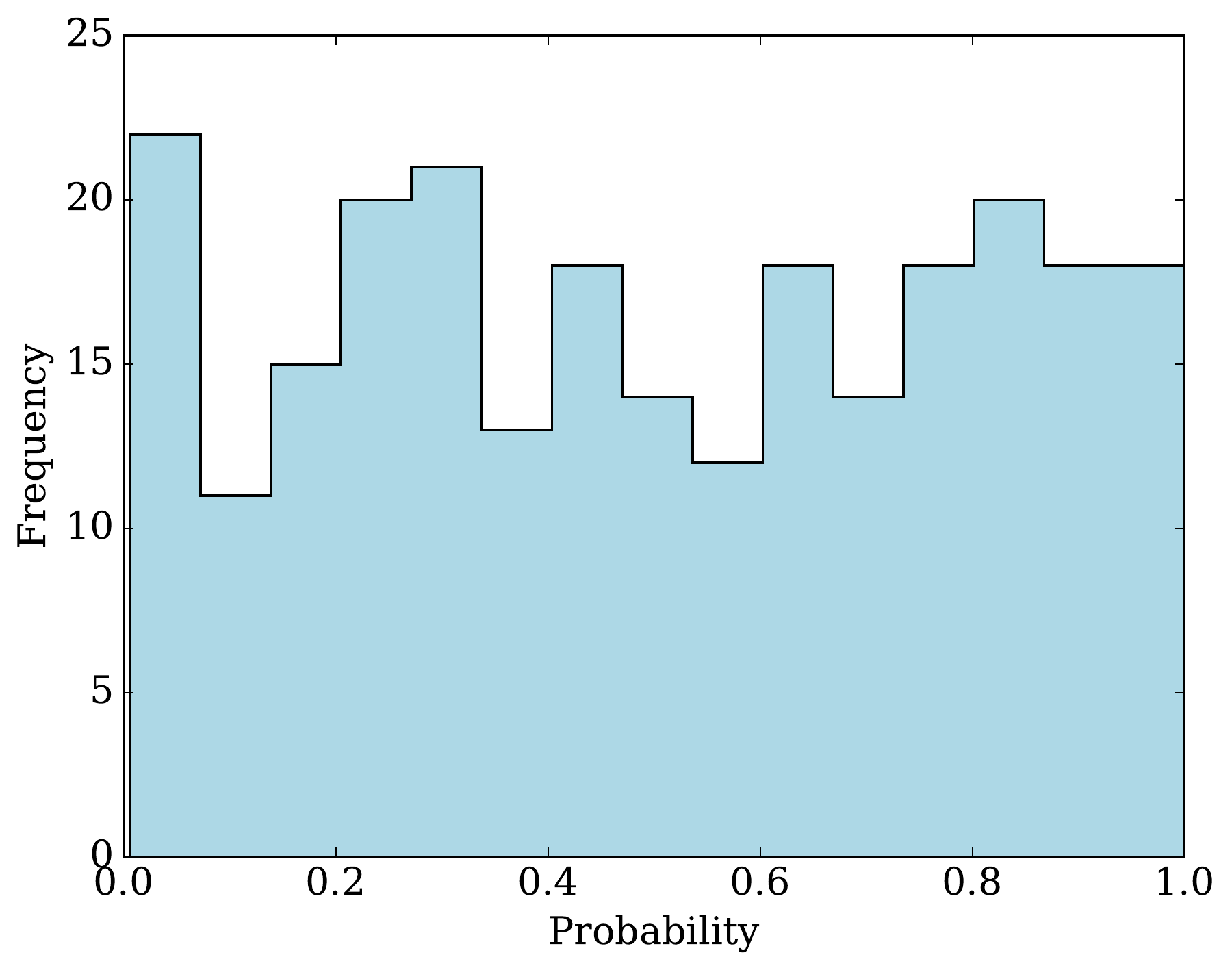}
\centering
\caption{Uniform distribution of the probabilities of the modelled PSDs for the 252 AGN from the MQ catalogue.}
\label{Fig:histprob}
\end{figure}

We have analysed 252 AGN using the {\it PSRESP} method to find the best-fitting power-law power spectral model of each system and derived the variability amplitude.  A short version of the catalogue of sources observed with {\it Kepler}/K2 is presented in a table in the Appendix (see Table \ref{Table:big}) and the full table can be found on-line. As explained in the previous section we used the de-trended light curves for the detailed analysis. However, we added in the table the power-law indices retrieved for the `optimal' light curves and the PRF9 light curves for reference. The table includes the general properties derived from the Million Quasar Catalogue (hereafter MQ) and the variability properties derived from this work. When cross-matching the observed targets with the latest version of the MQ catalogue we noticed that 23 objects did not appear in the current version. These sources were in the catalogue at the time of the proposal and they might have been removed because of their low false alarm probability. For that reason we have only considered the targets that are in the current version of the MQ catalogue for the detailed analysis, yielding a sample of 252 sources. We further discuss the possibility that some of the unidentified sources may also be AGN by exploring their variability properties. \\

\subsection{AGN of the Million Quasar Catalogue}

We present the results of the Monte Carlo simulations performed on the confirmed AGN to determine their best-fitting power-law PSD using the de-trended light curves. We  mostly used the `optimal' de-trended light curves but will also show in the table the `optimal' light curve and PRF9 light curves results for comparison. We used a simple model of a power-law with varying power-law index. An example of a fitted PSD is shown in Fig. \ref{Fig:psdexample} where the solid green line indicates the observed PSD and the black filled circles the simulated PSD. As explained in the previous section there is an artificial flattening at lower frequencies introduced by the Lomb-Scargle routine, but the first three bins were eliminated when comparing the observed and the simulated power spectra by means of the $\chi^{2}_{\rm dist}$. In the case of the PSD shown in Fig. \ref{Fig:prob} the power-law index is $-2.4\pm0.5$ and a goodness of fit of 83 \%.
After analysing all the targets in an analogous way we computed the probability distribution of the whole sample (see Fig. \ref{Fig:histprob}). The distribution of the p-values is uniform, this implies that all the data are consistent with a simple power-law model. For example, there are only 3 sources out of 252 with a p-value $< 1 \%$, which is exactly what we should expect if the simple power-law hypothesis is sufficient to explain the data for the entire sample. In the final catalogue that we present here we include all the fitted parameters, the acceptance probability and the observed and simulated fractional rms. It also contains the rest-frame simulated fractional rms, namely the variability amplitude corrected for time dilation, and its error. Additionally, the table has the basic parameters derived from the MQ catalogue such as the $b'$ and $r'$ magnitudes, the AGN type and the redshift. \\

\begin{figure}
\includegraphics[width=8.5cm]{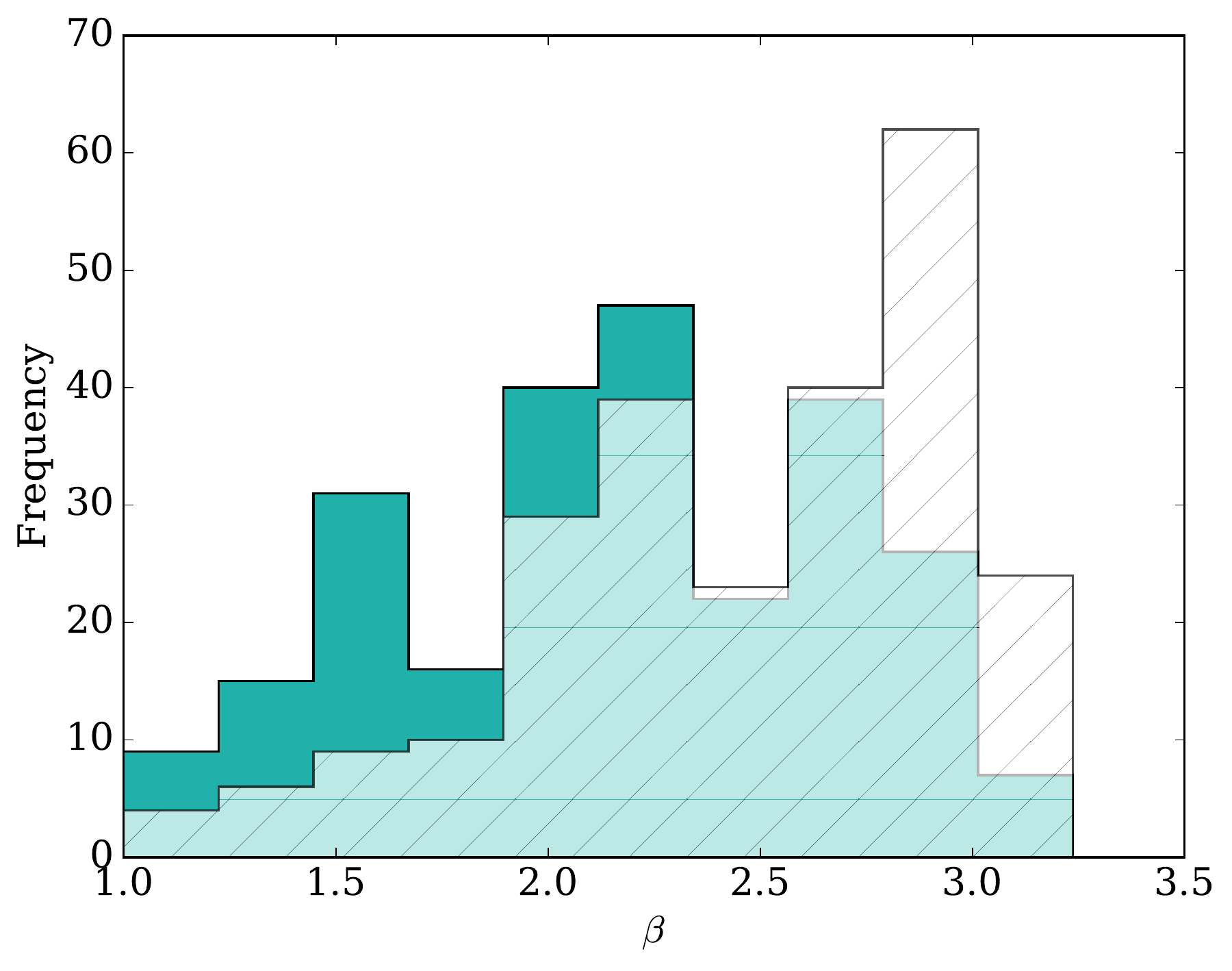}
\centering
\caption{Histogram of the power-law indices obtained using the `optimal' light curves in white and the `optimal' de-trended light curves in turquoise. The values of the two samples overlay and span from $-1.0$ to $-3.5$ with an average power-law slope of $-2.5$ and $-2.2$, respectively. }
\label{Fig:histbeta}
\end{figure}

We computed the distribution of power-law indices for the whole sample using the de-trended light curves, being the average $-2.2\pm0.5$ as shown turquoise in Fig. \ref{Fig:histbeta}. The distribution of power-law indices using the `optimal' light curves is also shown in white in Fig. \ref{Fig:histbeta}. It is skewed and peaks at around $-2.5\pm0.5$, steeper than the slopes derived using the de-trended light curves. This is similar to what have been shown in the previous section using artificial light curves. The variety of slopes found indicate that there is not a unique power-law index that describes the whole sample. There is no correlation observed between the probability and the parameter $\beta$, which indicates that there is no specific tendency for the best fitted cases to have a certain power-law index. 

\begin{figure}
\includegraphics[width=8.5cm]{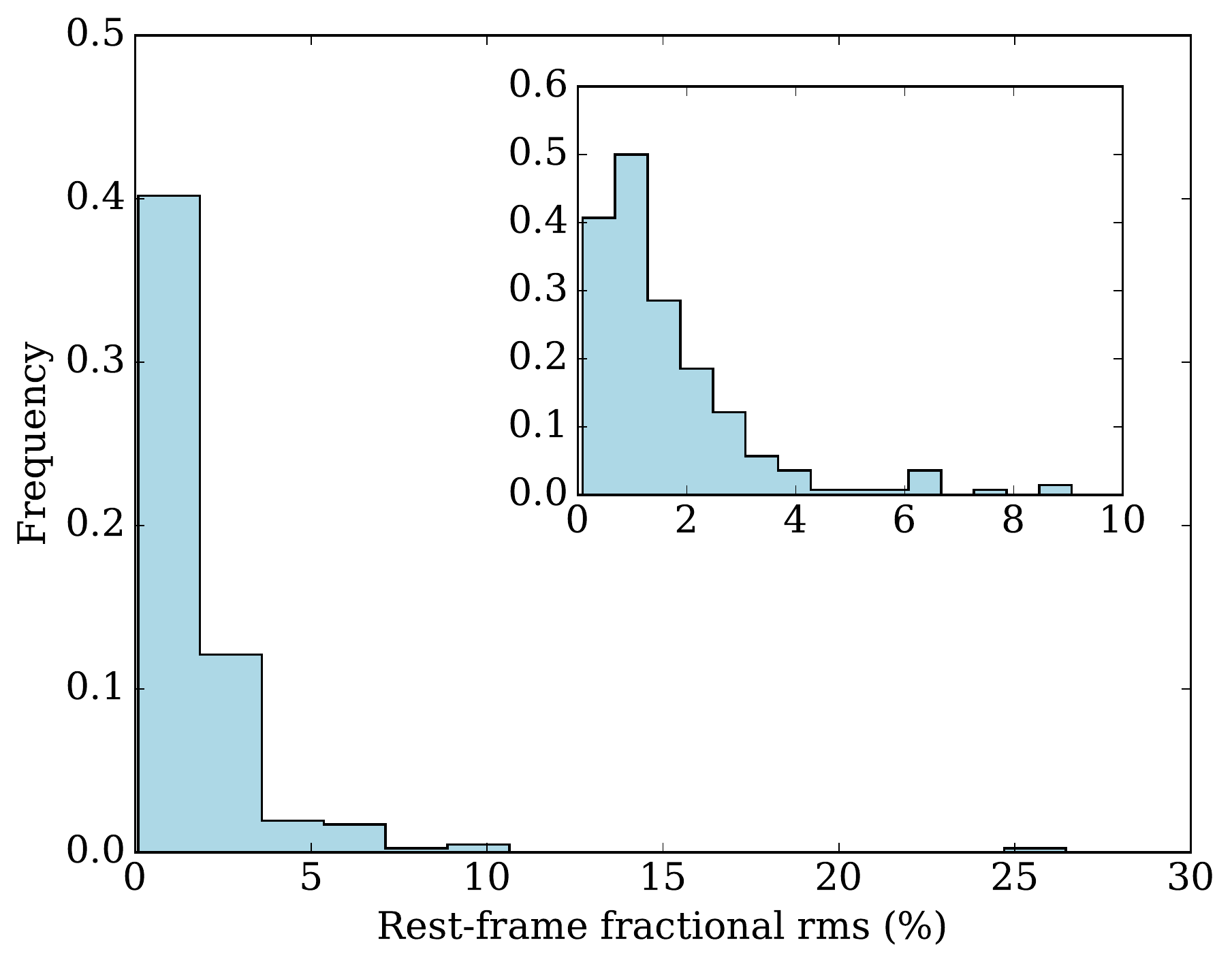}
\centering
\caption{Distribution of the rest-frame fractional rms. The quasar with a fractional rms of $\sim\,26\,\%$ corresponds to a blazar with EPIC number 201184312. A zoom on the histogram is provided in the upper panel, both are normed so that the integral over the range is 1.}
\label{Fig:histrms}
\end{figure}

We also present the distribution of the rest-frame fractional rms measured in the simulated power spectra in Fig. \ref{Fig:histrms}. The mean rest-frame fractional rms is 1.7 \%. There is one case EPIC 201184312 with a large value of amplitude variability of 26 \%, showing a fast strong variability with multiple flares, which is the typical behaviour of a blazar \citep{Healey2008}. \

\begin{figure*}
\includegraphics[width=7.5cm]{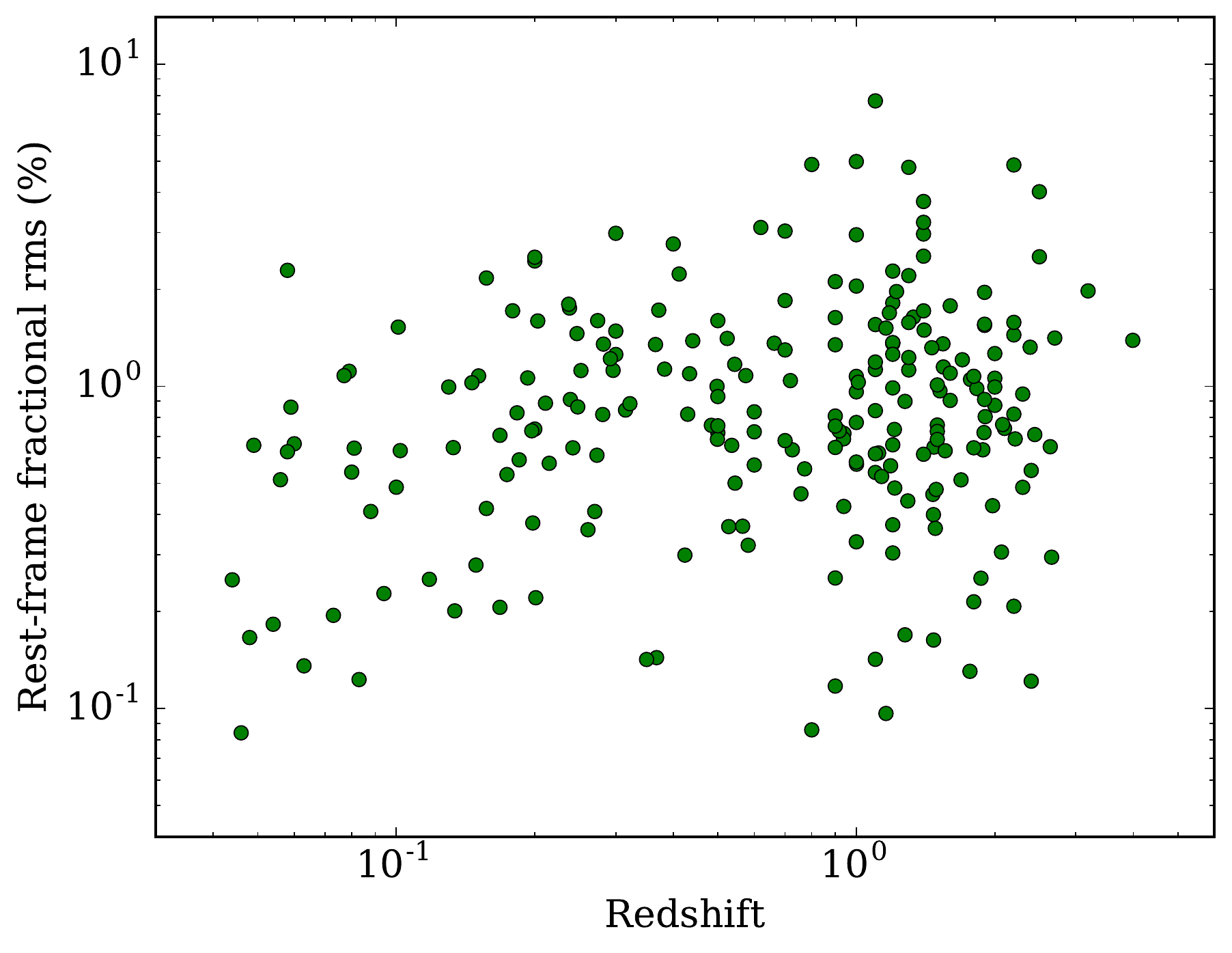}
\includegraphics[width=7.6cm]{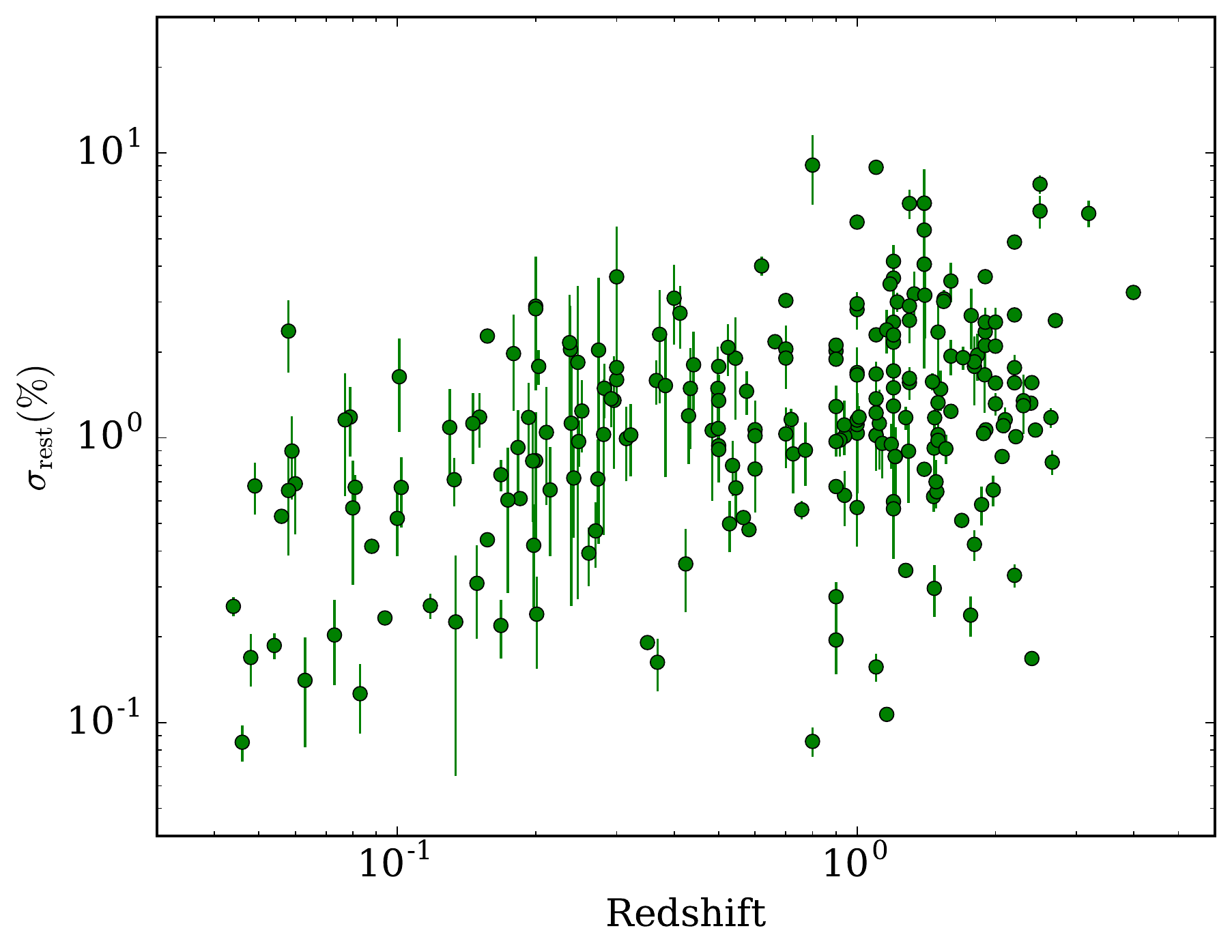}
\centering
\caption{{\it Left}: Simulated fractional rms versus the redshift. {\it Right}: Correlation between the rest-frame fractional rms and the redshift for the AGN with known redshift and excluding the blazar.}
\label{Fig:fracvsz}
\end{figure*}

In addition, we explored whether there is a correlation between the amplitude of variability and the redshift, using both the simulated fractional rms that we observed and the rest-frame fractional rms (see Fig. \ref{Fig:fracvsz}). To generate these figures we have excluded the cases with a large fractional rms of $>20$\,\% (to exclude the blazar) and we only used those AGN in our sample with known redshift. There is an apparent correlation between the rest-frame amplitude of variability and the redshift, thus we performed a Kendall rank correlation test to check if there is a monotonic correlation. Being the null hypothesis the absence of association between the two variables we have to reject it as the p-value is $3\times10^{-8}$ and the tau statistic is $\tau = 0.24$. According to this it might be a highly significant trend in the data, so that sources at higher redshift show larger variability, we further discuss the different physical explanations in Sect 5. Moreover, we do not find a significant correlation between the measured fractional rms and the redshift as the p-value is $0.02$ and the tau statistic is $\tau = 0.1$. \\

\begin{figure}
\includegraphics[width=8.5cm]{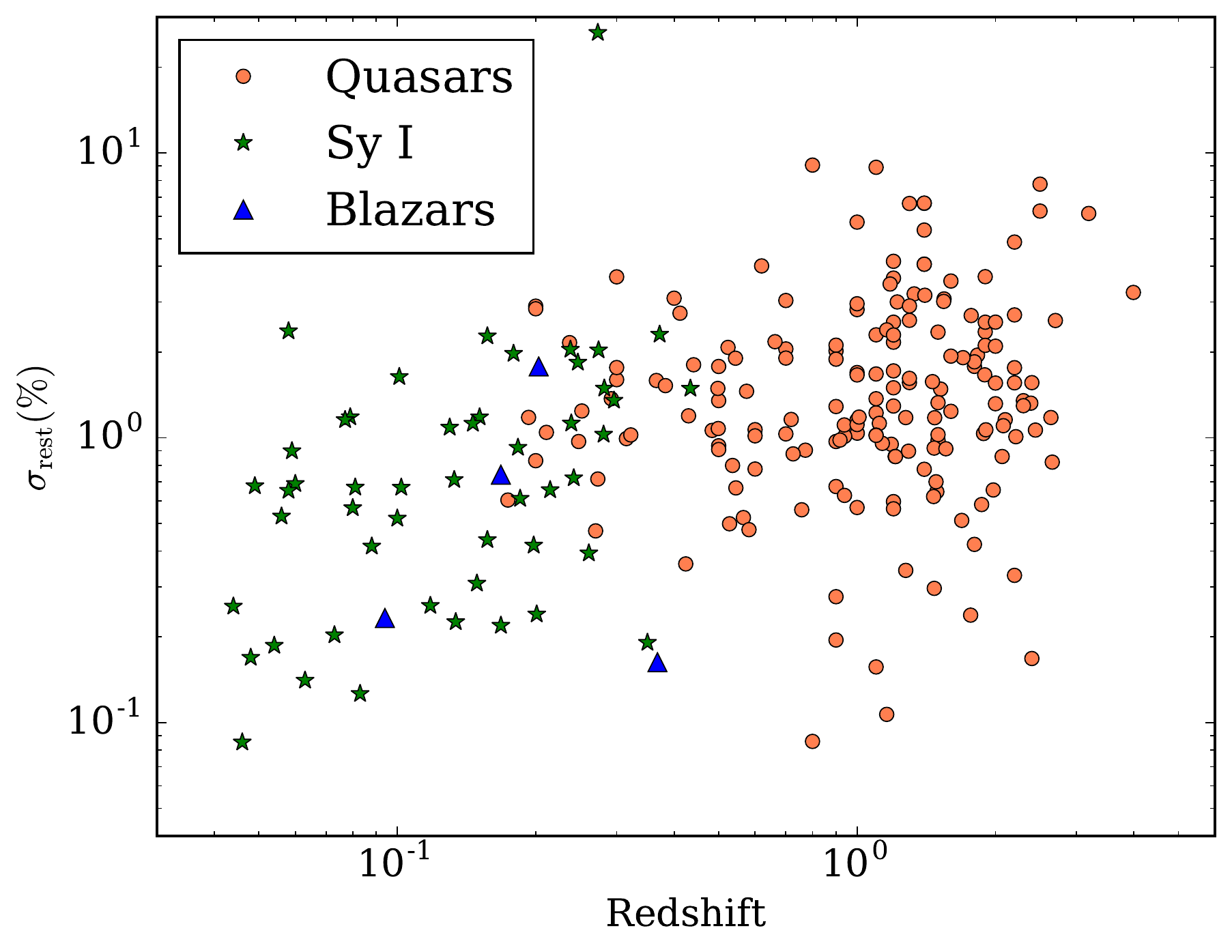}
\centering
\caption{Correlation between the rest-frame fractional rms and the redshift for the AGN with redshift for different types of AGN in the sample.}
\label{Fig:fracvsztype}
\end{figure}

We have also explored whether the different sub-classes of AGN in our sample show different variability properties. The sub-classes were extracted from the MQ catalogue and are indicated in the table shown in the Appendix. We have made three categories, 50 Seyfert I galaxies, 182 quasars and 5 blazars. There is only one Seyfert 2 type in our sample and 14 sources are unclassified. There is not a significant difference in the power-law index of the different categories. The rest-frame amplitude of variability shows that the Seyferts are indeed at lower redshift and show lower variability than the bulk of the quasars (see Fig. \ref{Fig:fracvsztype}). The sources classified as blazars do not present larger variability compared to the rest of the sample. The source with an amplitude of variability of 26\% is classified as a Seyfert I in the MQ catalogue but in other catalogues it is a blazar. The strong short-time variability derived in this work suggests that this source is indeed a blazar. 

\subsection{Unidentified sources}

We have also analysed the sources that were extracted from the MQ catalogue for the proposal but were removed in the current version of the catalogue. The latter might be explained by their low probability or absence of spectroscopic confirmation. However, some of these objects can be identified as AGN if they share the same variability properties as the MQ sample. We present a list of 4 sources that we think might be AGN since their PSDs show clear broad-band aperiodic noise and the average amplitude of variability is $1.3$ \%, of the same order as the MQ sample, and this is also confirmed by direct inspection of the light curves. For these sources the fit describes well the observed power spectra with an acceptance probability $>50$\,\%. They are listed in Table \ref{Table:notMQ} with their coordinates, fitted power-law index, the acceptance probability, the fractional rms and the rest-frame fractional rms. Since the redshifts are unknown we used the median of the whole sample $0.918$ to estimate the rest-frame amplitude variability. There was not a coincident source for the first three objects listed in Table \ref{Table:notMQ} in any of the consulted catalogues, therefore they are unclassified or new sources. However, the EPIC 206072629 can be associated with the X-ray source 1RXS J$215935.6-124859$, for which there is no more information in the literature. \\

\begin{table*}
\footnotesize
 \centering
 \caption{Sources not identified in the MQ Catalogue. The table includes the EPIC number, the best negative power-law index $\beta$ for the `optimal', PRF9 and de-trended light curves. For the de-trended cases the table shows the acceptance probability, the observed and fitted fractional rms, and the fractional rms corrected from time dilation using as redshift the median of the MQ sample.}
\begin{tabular}{cccccccccc}
\hline\hline
EPIC & RA & DEC &  $\beta_{\rm opt}$  &  $\beta_{\rm PRF9}$  & $\beta_{\rm det}$  & Probability (\%) &  $\sigma_{\rm obs}$ (\%) & $\sigma_{\rm fit}$ (\%)  & $\sigma_{\rm rest,fit}$ (\%) \\
\hline
\input{latextable_14novnon}
\hline
\end{tabular}
\label{Table:notMQ}
\end{table*}
\section{Discussion}

We have analysed the light curves of the largest sample of AGN (252) to date observed with {\it Kepler}/K2 at cadences of $\sim30\,\rm min$ and generated a catalogue with all the variability properties derived from the study of their PSDs (see Table \ref{Table:big} and the full table on-line). The PSDs were fitted using the Monte Carlo technique assuming a simple power-law model that satisfactorily fit all of the AGN in our sample. Hence, breaks and QPOs are not required to describe the PSDs of AGN observed with {\it Kepler}/K2. We have also calculated the fractional rms and explored the correlation with the redshift. 

\subsection{Systematic effects}

K2 light curves suffer from instrumental signals. The most relevant systematics for our study are the {\it differential velocity aberration} effect and the Moir\'e pattern that is seen in some targets. There are 12 AGN in our catalogue whose light curves might be affected by the Moir\'e effect and the rolling band and are indicated with a star in the catalogue. Even though these sources do not present different properties compared to the non-affected ones, they must be treated with caution. We have extensively studied the effect of the DVA effect, and found that de-trending the light curves with a sine curve at the orbital period of the satellite is the best approach to mitigate the long term instrumental trends. This does certainly not remove all systematics, which is a clear disadvantage of using {\it Kepler}, which was designed to be stable on typical time-scales of less than a day. Using larger apertures also minimizes the effect but the associated increase in noise limits the use of these data. 

\subsection{Comparison with previous variability studies}

In this work a simple power-law model adequately describes the PSDs of our sample, but we cannot rule out the presence of a frequency break at lower frequencies in the power spectra. The high frequency break in the PSDs is expected to be at about a year time-scale. For the analysis we split the light curve in segments and then determined the Lomb-Scargle periodogram, so that the lowest frequency that we can reach is $>10^{-6}$ Hz. \citet{Mushotzky2011} report similar results, they fitted the PSDs of four {\it Kepler} AGN with a power-law model. However, it is known from previous variability studies in the X-rays that a break to a slope of $-1$ is present  \citep[see][]{Uttley2002,Markowitz2003,GonzalezMartin2012,Marshall2015}. The latter seems to correlate with the mass of the black hole, so that the more massive the black hole is the longer is the characteristic time-scale \citep[see][]{McHardy2006}. Likewise, using {\it Kepler} observations \citet{Shaya2015} report similar breaks at $\sim1\times10^{-6}$ Hz for a large sample of AGN. Moreover, in recent studies of the AGN Zw 229--15 they found a break at $2.3\times10^{-6}$ Hz where the slope changes from $-2$ to $-4$ \citep{Carini2012,Edelson2014}. This 5 day characteristic time-scale is not consistent with the viscous time-scale as it should be of the order of years for that object. But there should be a break at much longer time-scales not covered by {\it Kepler} similar to what has been found in other studies on variability. According to the work presented here there is not such a break on fast time-scales, but further investigation is needed to constrain the optical PSD shapes of AGN in the high frequency domain to better understand the physical processes and the origin of the fast variability observed. The X-ray emission is thought to originate very close to the black hole in the corona, whereas the optical emission is coming from the accretion disc itself, both due to intrinsic disc emission as well as reprocessed light due to X-ray heating of the disc. \\

The average measured power-law index of $-2.2$ reported in this work is slightly steeper than the slope of $-2$ found in the X-rays, moreover we find a wide range of power-law slopes from $-1$ to $-3.2$ \citep[see][for the X-ray PSD slopes]{Uttley2002,GonzalezMartin2012}. Power-law slopes of $-2$ are consistent with the the {\it damped random walk} (DRW) model, one of the most commonly used models to study optical variability \citep[e.g.][]{Kelly2009}. Similar results as in X-rays have been reported in optical variability studies on longer time-scales than studied here using ground-based surveys such as the Palomar-Green quasars \citep{Giveon1999Palomar}, {\it MACHO} \citep{Kelly2009} and SDSS Stripe 82 \citep{MacLeod2010}. However, the DRW model fails to capture the behaviour exhibited by AGN observed with {\it Kepler} as demonstrated by \citet{Kasliwal2015DRW}. Recent studies on {\it Kepler} AGN report steep PSDs \citep[see][]{Shaya2015,Mushotzky2011,Edelson2014} but also in an even more recent study of AGN using the PTF/IPT survey \citep{Caplar2017}. The steep slopes observed could be attributed to viscous damping of high frequency accretion fluctuations combined with the filtering effect of the extended disc emission region responsible for the emission in the {\it Kepler} bandpass \citep[e.g. see][for a discussion of the effects of an extended emission region on light curves driven by accretion fluctuations]{Arevalo2006}. Steep PSDs are also in better agreement with magnetohydrodynamics simulations (MHD) of accretion discs \citep[e.g.][]{Reynolds2009a}. The variety of slopes from $-1$ to $-3.2$ found here are consistent with previous results described by the DRW model but also with recent studies reporting steeper slopes.  \\

\subsection{Correlations with physical parameters}

We explored the correlations between the amplitude of variability and parameters of the system such as the redshift and the bolometric luminosity. In Fig. \ref{Fig:fracvsz} we presented a correlation between the rest-frame amplitude of variability and the redshift, so that the variability seems to be larger at higher redshift. This correlation has been reported in previous studies \citep[e.g.][]{Giveon1999Palomar,vandenBerk2004}, but not in recent variability studies of AGN such as the PTF/IPTF or the PanSTARRS1 survey \citep[see respectively][]{Simm2015,Caplar2017}. \

We consider it unlikely that there is a cosmological evolution of the variability amplitude, since AGN of a given luminosity do not show any evidence for evolution in their spectral energy distribution \citep[][]{Steffen2006}, indicating that the central engines are the same regardless of redshift. The most likely explanation for the correlation between the rest-frame amplitude of variability and the redshift is that it is caused by the known ``bluer when brighter" effect in quasar variability, arising from the fact that emission at shorter wavelengths is more variable \citep[e.g.][]{diClemente1996,Cristiani1997,Giveon1999Palomar,vandenBerk2004}. \citet{Sun2014} show that the effect is time-scale dependent and even stronger on the short time-scales we consider here. For example, on time-scales of 20 days, Sun's low redshift sample shows a 20~per~cent increase in variability amplitude from SDSS $r$ to $g$ bands, while \citet{Zhu2016} use similar methods with GALEX data to show a 50~per~cent increase in variability amplitude from NUV to FUV. The wavelength differences corresponding to these changes in variability amplitude are relatively small ($\sim30$~per~cent or less), compared to our redshift range of $z=0$ to $\sim 2$, leading us to expect even larger systematic changes in variability amplitude as we observe sources from low to high redshift.\

\begin{figure}
\includegraphics[width=8.5cm]{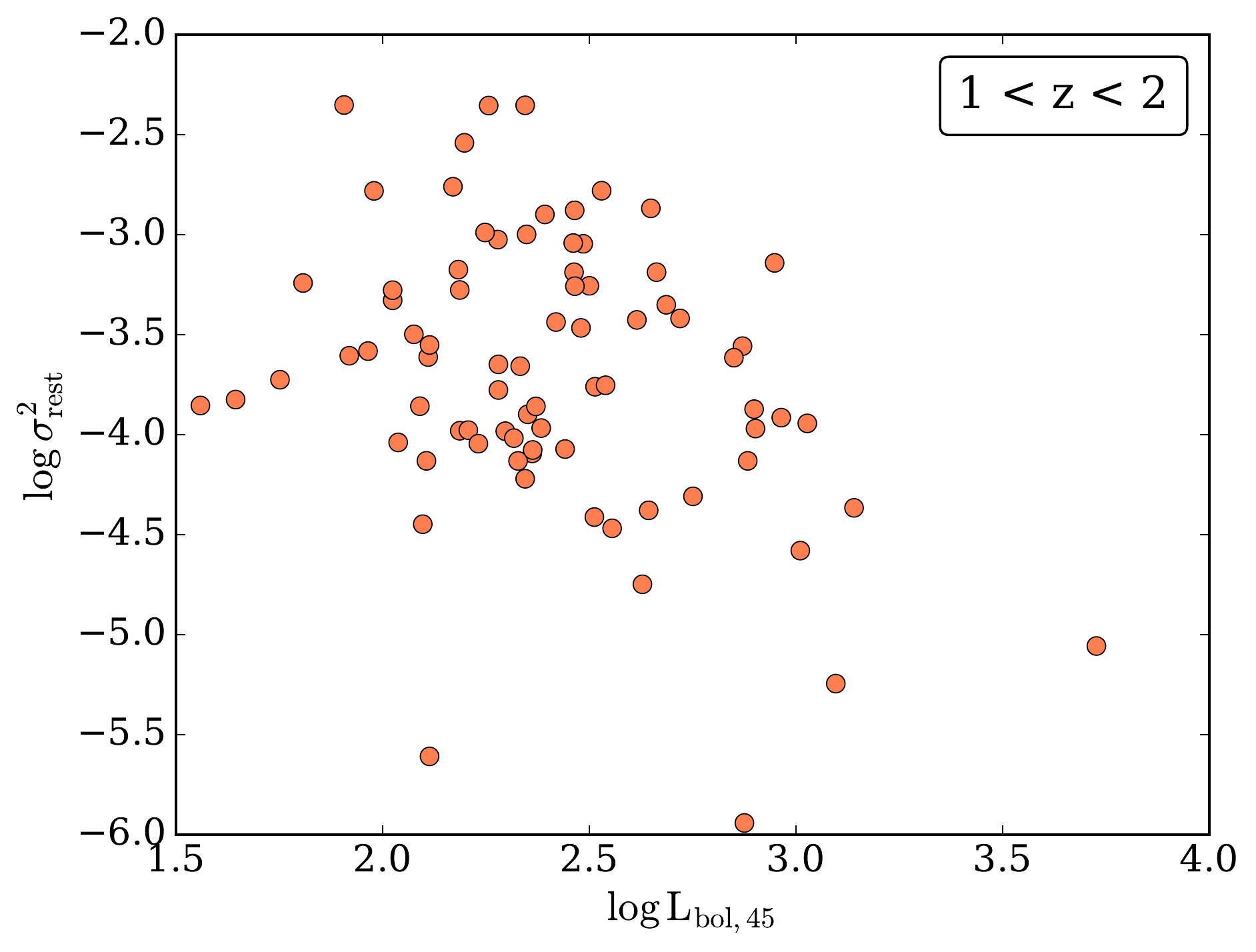}
\centering
\caption{Correlation between the rest-frame fractional rms and the bolometric luminosity for the AGN with redshift $1 < z < 2$ and excluding the blazar to generate this figure.}
\label{Fig:fracvslum}
\end{figure}

\begin{figure}
\centering
\includegraphics[width=8.5cm]{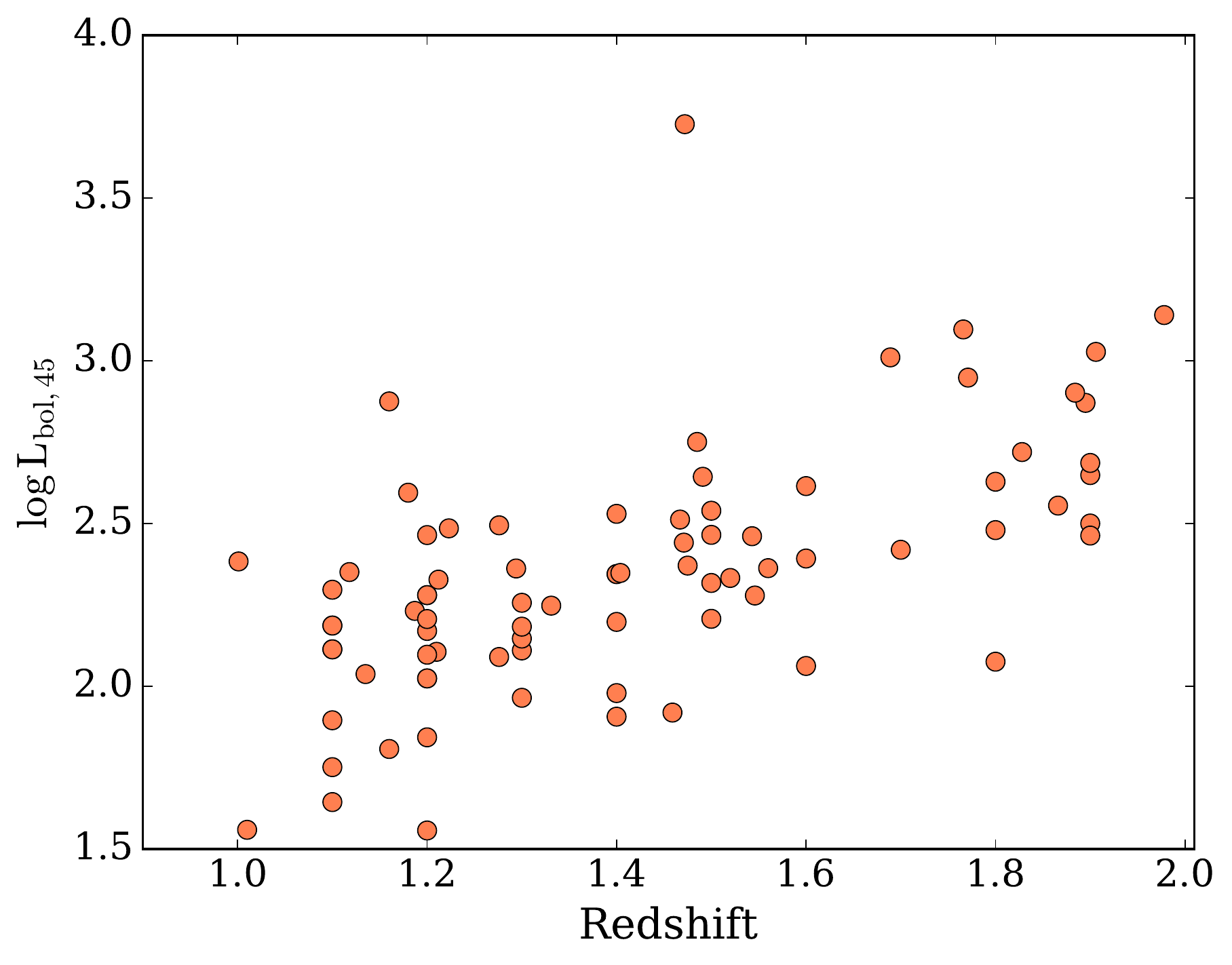}
\caption{Correlation between the calculated bolometric luminosity of the AGN with known redshift with a p-value $\sim\,2\times10^{-9}$ and $\tau$-statistic $0.5$.}
\label{Fig:lumvsz}
\end{figure}

Recently, \citet{Simm2015} report an anti-correlation between the amplitude of variability and the bolometric luminosity on month and year time-scales for AGN at $1 < z < 2$. The latter is a good proxy for the mass accretion rate of the system, so that higher accreting objects show lower variability. Motivated by this result we have searched for the same relation on hour time-scales using the MQ sample. We estimated the bolometric luminosities by using the apparent magnitudes in the b$'$ band extracted from the MQ catalogue that were extracted from POSS-I. We converted them into absolute magnitudes using the K-corrections from \citet{Kcorrection} in the B band (only up to $z\sim2.2$) and then we obtained the luminosities in this band. To convert the luminosity in the B band into the bolometric luminosity we used the relation described in \citet{Hopkins2007bolometric}. We have included the estimated bolometric luminosities in the Table \ref{Table:big}. There is tentative correlation between the variability amplitude on time-scale of hours and the estimated bolometric luminosity. According to the Kendall tau test, the p-value $0.03$ and $\tau$-statistic $-0.17$ might be an evidence of the anti-correlation, but it is very marginal (see Fig. \ref{Fig:fracvslum}). Here we consider significantly shorter time-scales than studied by \citet{Simm2015}, so it is possible that the anti-correlation between luminosity and variability amplitude does not apply on short time-scales. However, our sample also shows a significant correlation between luminosity and redshift (Fig. \ref{Fig:lumvsz}) which is expected due to both the Malmquist bias and the known cosmological evolution of the quasar luminosity function \citep{Boyle2000}. Therefore, we consider it likely that any anti-correlation between luminosity and amplitude is obscured by our observed correlation between variability amplitude and redshift. Furthermore, we did not find a correlation between the slope of the PSDs and the bolometric luminosity. A more detailed study of AGN with a wider range of measured luminosities is required to study how the variability is associated to both the luminosity and the black hole mass.\\

\subsection{Source identification}

Besides, by studying the fractional rms we were able to identify a blazar as seen in Fig. \ref{Fig:histrms}. We observed an AGN with extremely high variability amplitude of  $\sim26\,\%$ compared to the majority of the sample that have a fractional rms of less than $4\,\%$. This object with EPIC number 201184312 (2dFGRS TGN172Z225) is a blazar at redshift 0.27. Blazars present stronger variability than radio-quiet quasars and the variable emission is thought to be produced from the shocks in the jet, and not in the accretion flow. Furthermore, this source has associated radio emission according to the FIRST Survey at 1.4 GHz with a flux of 135 mJy \citep{BeckerFIRST}. It also appears in the {\it Roma}-BZCAT catalogue of blazars classified as a flat-spectrum radio quasar \citep[e.g.][]{Massaroblazars2015} and has a $\gamma$-ray counterpart detected with {\it Fermi/LAT}. We could identify another case with strong fast variability in the sample, the object with EPIC 201600065 presents a very strong flare 12 \% brighter than the persistent level with a fast rise of order of 3 days and a duration of $\sim8$ days. The rest-frame fractional rms is similar to the rest of the sample  $\approx 2\% $. The shape and the time-scales of this rare event can not be associated with a tidal disruption event as these have a duration of months and a rise time of weeks \citep{HungTDE2017}. The object is known as PKS 1106+023, it is classified as a Seyfert 1, it is nearby at redshift $0.157$, the black hole mass is $\sim 5\times10^{7}\,\rm M_{\odot}$ and its radio loudness is $\sim 2\times10^{3}$ \citep{PKSsource}. There is a clear radio jet feature in a FIRST image at 1.4 GHz \citep{BeckerFIRST1995}, so it might be that the angle respect to the observer is such that the optical variable emission is coming from both the jet and the accretion flow. Besides classifying AGN it is also possible to confirm AGN candidates by looking at their variability properties as it has been demonstrated in Sect. 4.2. We believe that at least four sources that were excluded from the MQ catalogue are AGN by looking at their light curves, PSDs shape and fractional rms.  Hence, the study of fast variability properties of a big sample of AGN is a very powerful tool to identify AGN. \\

\section{Conclusions}

In this work, we presented a catalogue of the optical variability properties in the high-frequency domain of a large sample of 252 sources. Using light curves of extremely high-quality from {\it Kepler}/K2 we studied their power spectra by using the {\it PSRESP} Monte Carlo method. Our conclusions are summarised as follows.\

\begin{enumerate}
\item {\it Kepler}/K2 power spectral densities of AGN are well described by a simple power-law in a frequency range $6\times10^{-6}-10^{-4}$ Hz. A break at lower frequencies $\approx10^{-6}$ Hz is not required and it is expected on longer time-scales. 
\item We found a variety of power-law slopes suggesting that all the light curves cannot be simply modelled with the {\it DRW} model. This is supported by other authors using {\it Kepler} observations that reported steeper slopes.
\item The typical variability of the sample is $\sim 2\%$ in the frequency range of $6\times10^{-6}-10^{-4}$ Hz, but we found a wide range of amplitude of variability ranging from $0.1-26$ \%. 
\item There is a significant correlation between the rest-frame fractional rms and the redshift that we believe is associated with the wavelength dependence of the variability. Thus, the amplitude of variability emitted originally in the UV is larger but we observe it in the optical wavelengths together with objects that are nearby and therefore present lower variability.
\item The fast time-scales explored here do not show the clear anti-correlation between rest-frame fractional rms and the bolometric luminosity seen in other samples on longer time-scales. It might be a slight anti-correlation, but it is likely that the expected anti-correlation is obscured in our sample, because we preferentially see higher luminosity sources at higher redshifts, which also show systematically higher rest-frame rms.
\item These type of optical variability studies on short time-scales are excellent to identify blazars, as they generally show much higher variability. Furthermore, we were able to identify 4 AGN that were dropped out from the MQ catalogue by comparing their variability properties with the rest of the sample.
\item Instrumental trends can significantly impact the measured spectral density power-law index. We find that de-trending the light curve using a single sine curve at the orbital period of the spacecraft is adequate for our study but it does not remove all instrumental signals.
 \end{enumerate}
 
\section*{Acknowledgements}

This work is partly financed by the Netherlands Organisation for Scientific Research (NWO), through the VIDI research program Nr. 639.042.218. We would like to thank Andrew Vanderburg for his advice regarding the systematic effects of K2 on the light curves. We would also like to thank the anonymous referee for the useful remarks and comments that allowed us to improve the paper significantly. \\



\bibliographystyle{mn2e}
\bibliography{rms4} 


\appendix

\section{Effect of a trend in a light curve on the power spectrum}
\label{Ap:trendeffect}

\subsection{Simulated data}
\begin{figure}
\centering
\includegraphics[width=6.5cm]{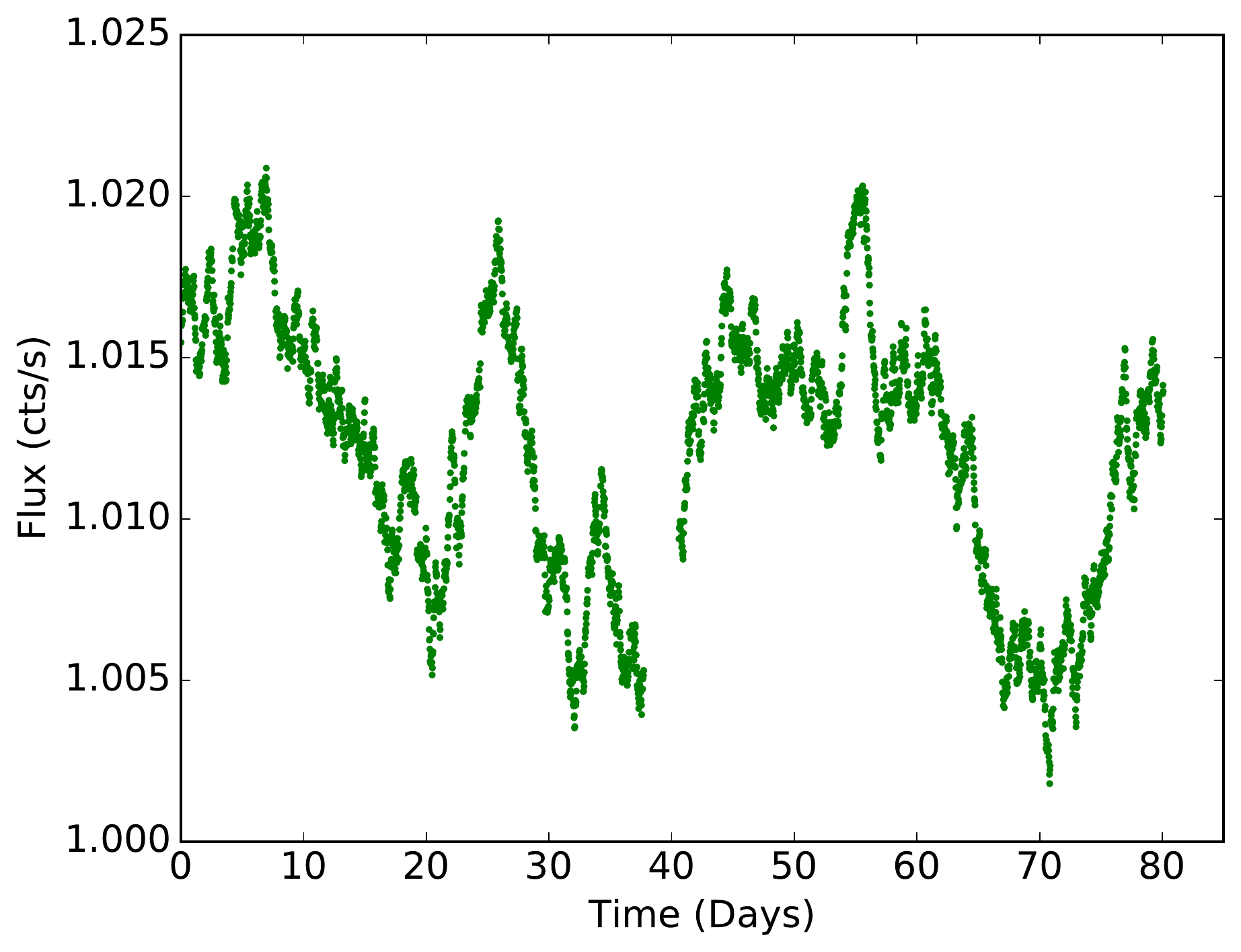}
\includegraphics[width=6.5cm]{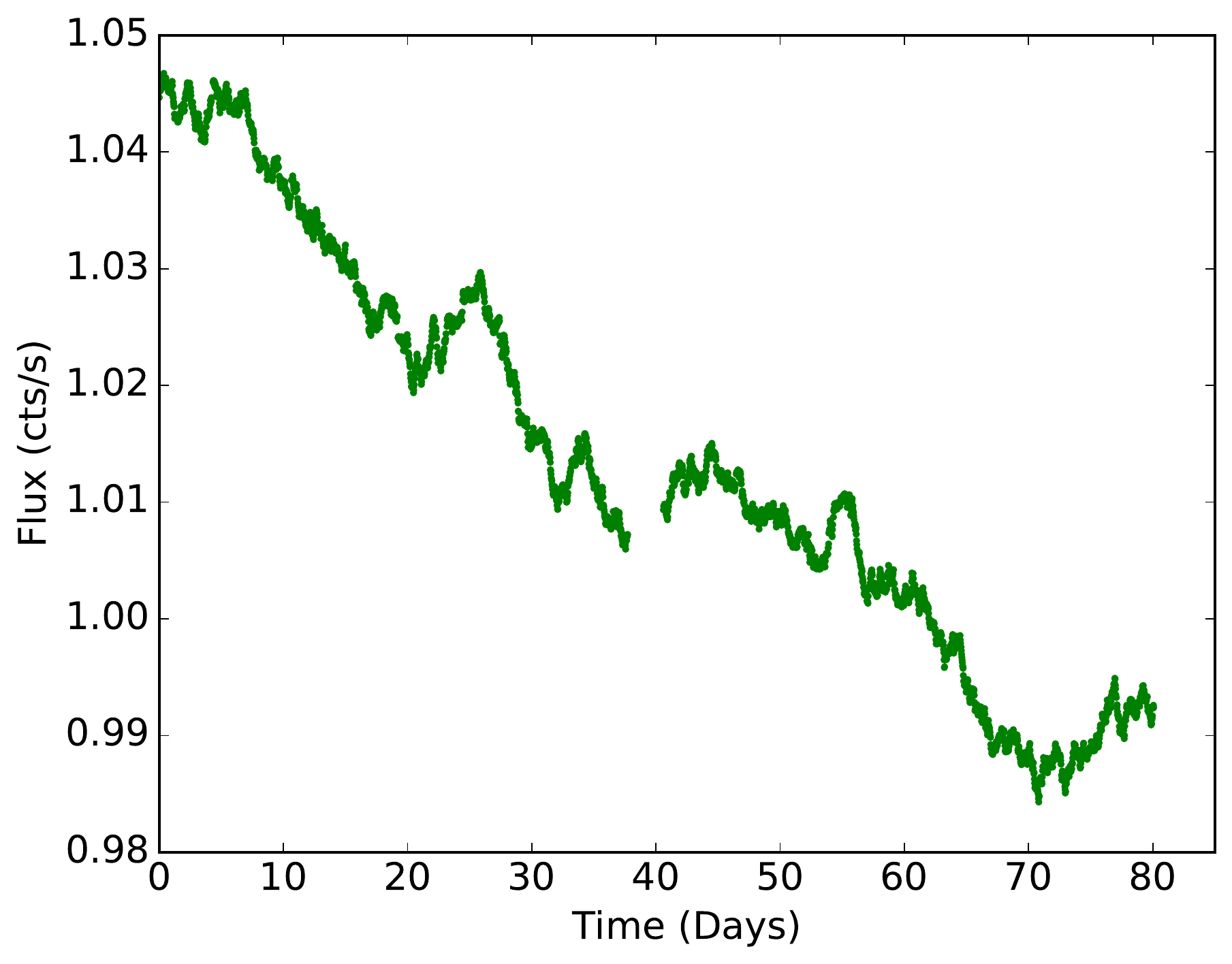}
\centering
\caption{{\it Upper}: Simulated light curve. {\it Lower}: Simulated light curve with a trend similar to K2 AGN light curves.}
\label{Fig:fakelcr}
\end{figure}

\begin{figure}
\centering
\includegraphics[width=6.5cm]{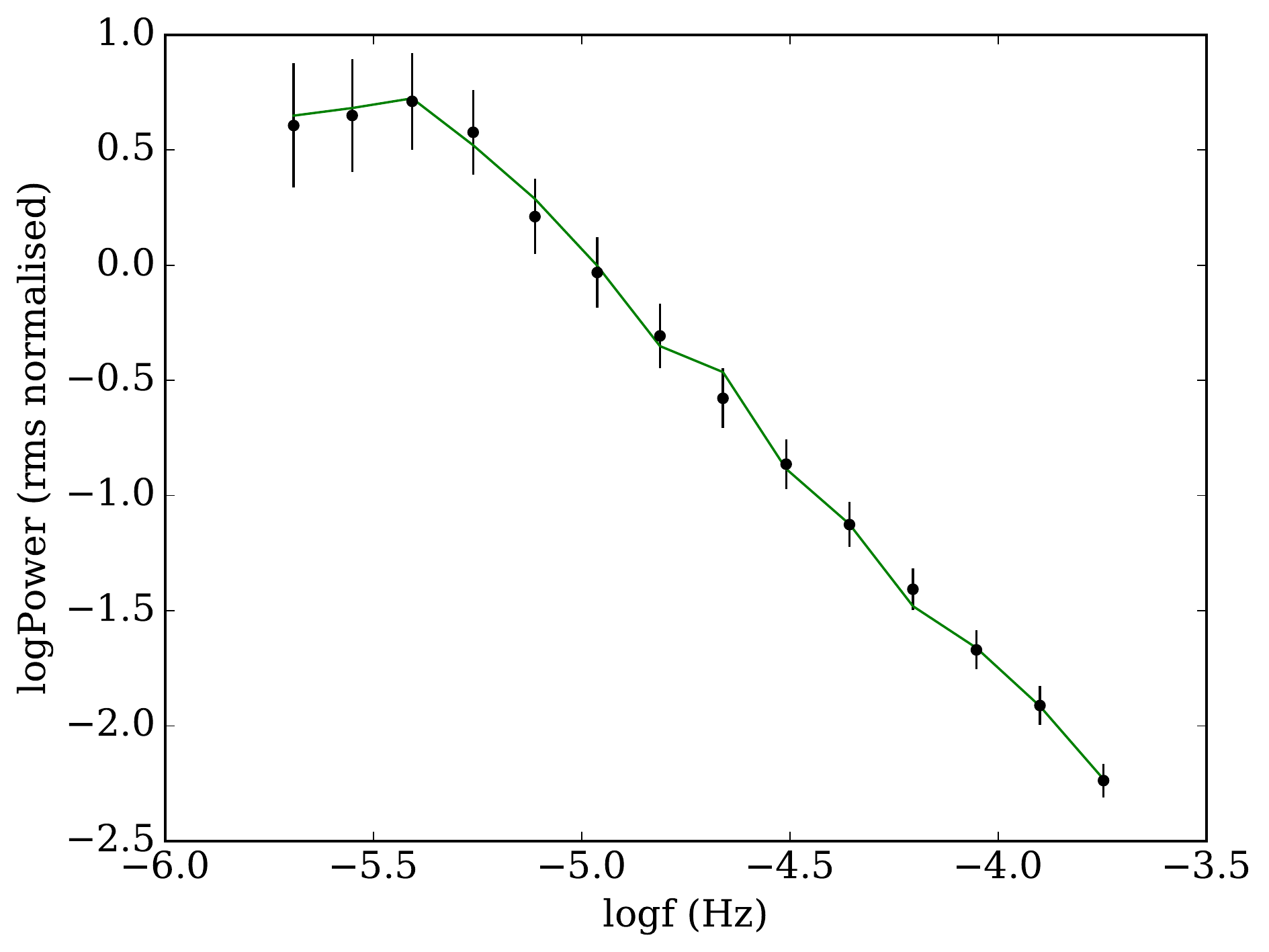}
\includegraphics[width=6.5cm]{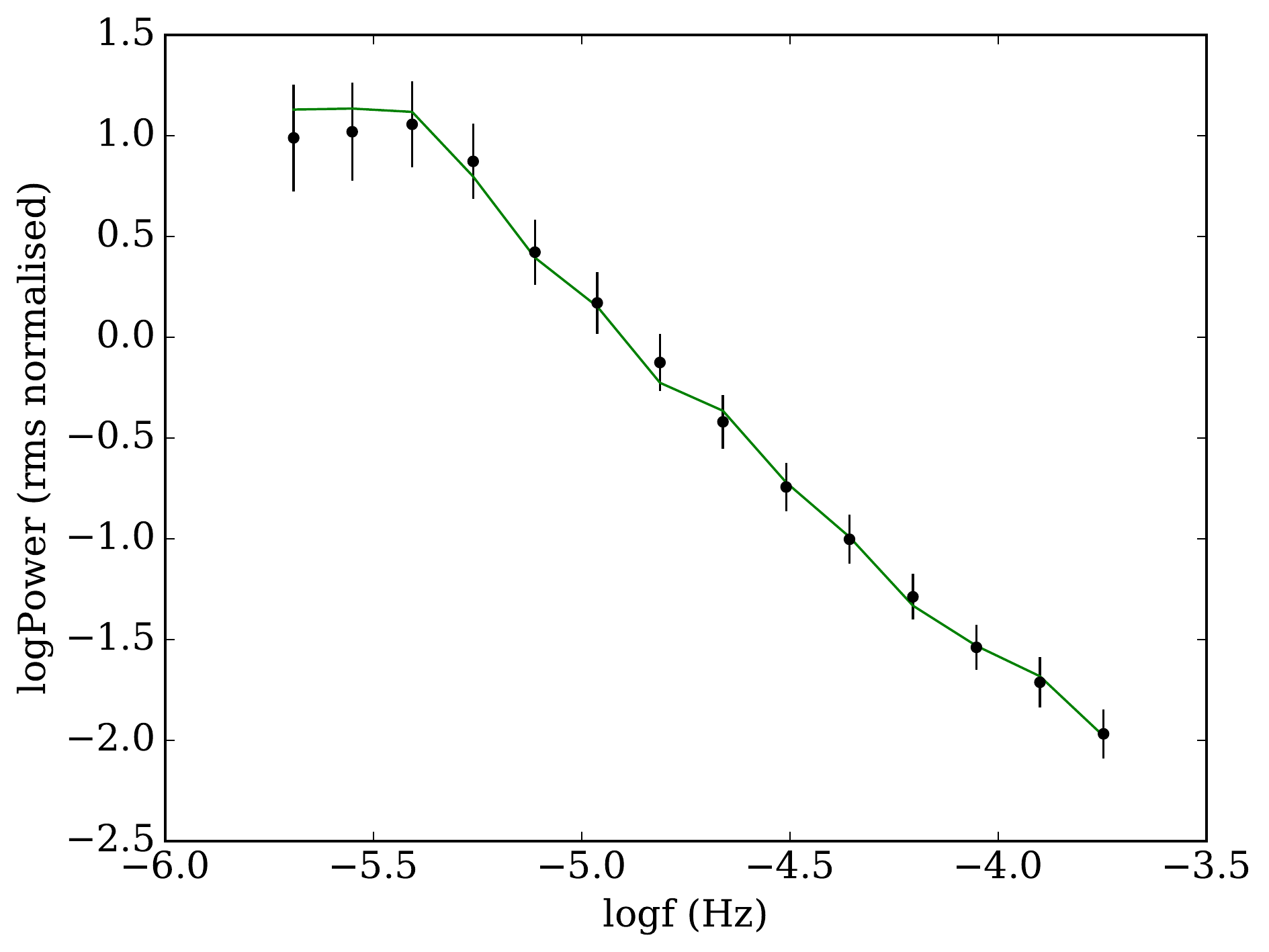}
\centering
\caption{{\it Upper}: Power spectra of the artificial light curve generated by \citet{Timmerkoenig}. {\it Lower}: Power spectra of the artificial light curve with a trend similar to the one seen in K2 AGN light curves, the power spectrum is steeper with $\beta\approx-3$.}
\label{Fig:fakePSD}
\end{figure}

\begin{figure}
\includegraphics[width=8cm]{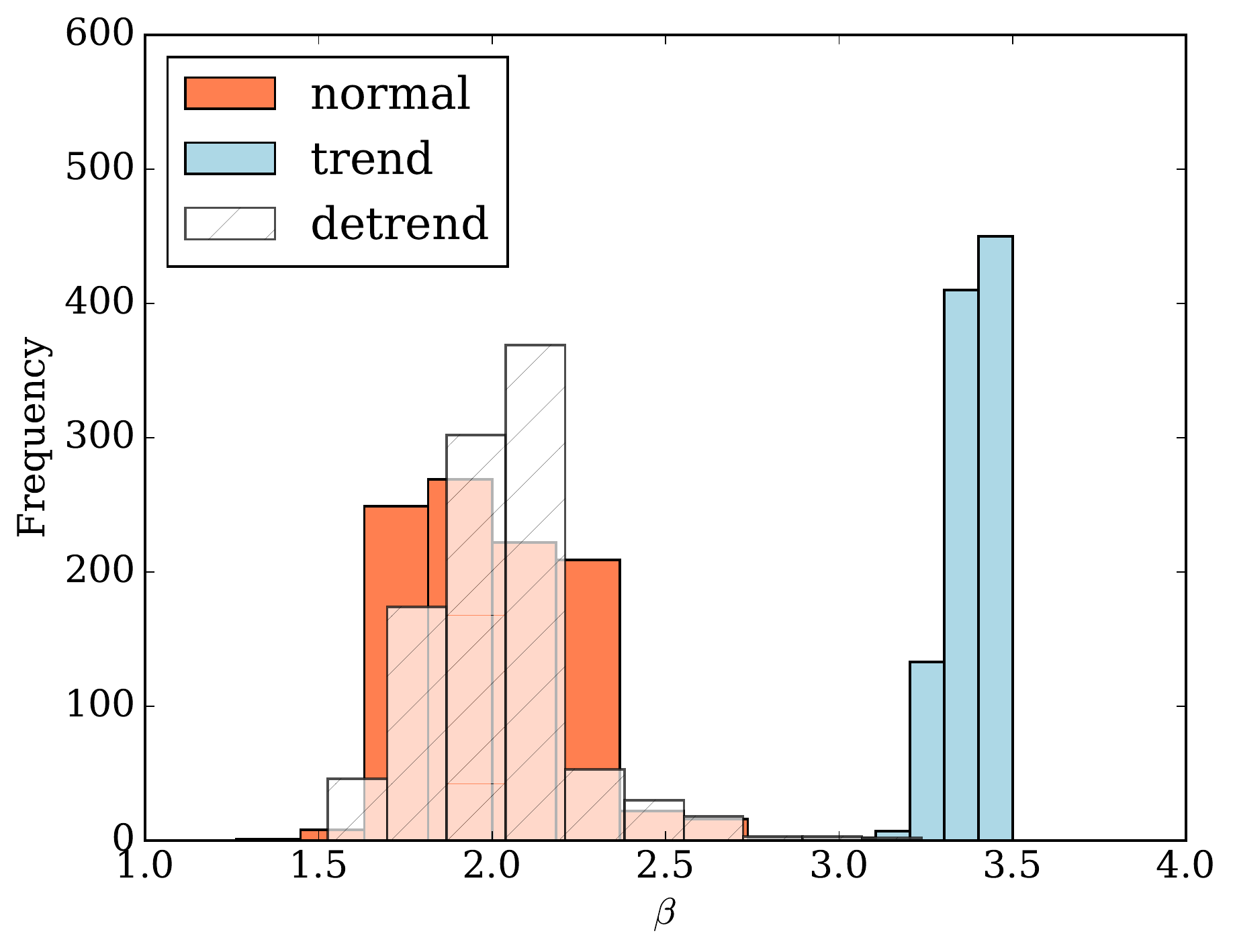}
\centering
\caption{This histogram illustrates the comparison between the best $\beta$ found via the {\it PSRESP} method for the 1000 original simulated light curves (in orange), the trended light curves (in light blue) and the de-trended (in white with stripes). }
\label{Fig:simulcom}
\end{figure}

We investigate whether trends in the light curves similar to the ones observed in the K2 light curves have a real effect on the slope and the normalisation of the power spectrum. For this we first simulate 1000 fake light curves using the \citet{Timmerkoenig} with a power-law model with a slope of $-2$ and we used the sampling pattern, the mean and the variance of the AGN 201187315 (see example in the upper panel of Fig. \ref{Fig:fakelcr}). Then, we analysed the artificial light curves using the PSRESP method and obtained the best model parameters (see upper panel of Fig. \ref{Fig:fakePSD} for an example). \
To determine a realistic trend, we fit a sine curve with the known period of the orbit of {\it Kepler} to the light curve of the AGN 201187315 and derived the best-fitting values for the phase and the amplitude. We multiplied our artificial light curves by this sine curve, thereby introducing a trend in the artificial light curve similar to what we observe in {\it Kepler} light curves (see example in the bottom panel of Fig. \ref{Fig:fakelcr}). Next, we analysed the trended artificial light curves with the PSRESP method (see example of the lower panel of Fig. \ref{Fig:fakePSD}). For both cases the acceptance probability was higher than $95\%$. We observe a steepening in the power spectrum in the trended light curves as the average for the 1000 light curves is $\beta_{2}=-3$. After de-trending the light curves, we find an average $\beta_{2}=-2$. It is important to note that even if there is a difference compared to the slope of the power spectrum of the normal light curve, the slope was within the statistical errors derived from the model. The histogram showing the different slopes for the normal artificial light curves and the trended ones is shown in Fig. \ref{Fig:simulcom}. In this figure we also include how the real power-law index is recovered when we de-trend the trended artificial light curves. 

\subsection{Real K2 data}
\begin{figure}
\includegraphics[width=8cm]{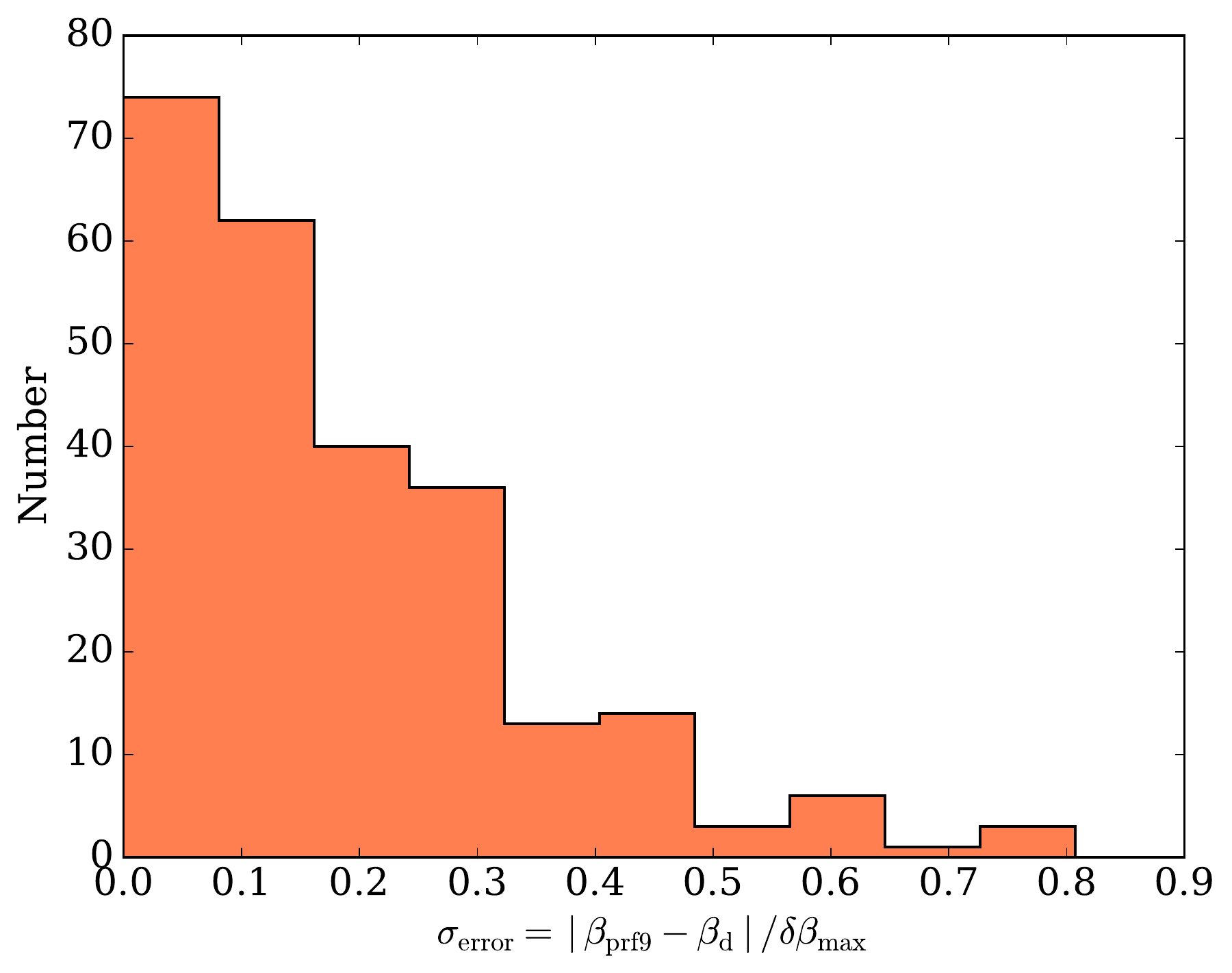}
\centering
\caption{This histogram illustrates the significance error between the slopes found using the light curves extracted from the PRF9 aperture and the de-trended light curves. This shows that the two sets of results are in agreement with each other.}
\label{Fig:significance}
\end{figure}

We have investigated the difference in the results provided by our PSRESP pipeline when using real K2 data. We have measured the power-law slopes using de-trended `optimal' light curves and light curves extracted with the largest aperture available in \citet{Vandenburg2014}, i.e.~PRF9. This largest aperture is the least affected by the {\it differential velocity aberration} effect. As shown in Fig. \ref{Fig:significance}, the difference in slopes measured from PRF9 and de-trended `optimal' light curves is well below $1\,\sigma$ in all cases. We therefore decided to use the de-trended `optimal' light curves, as these benefit from much lower Poisson noise than the PRF9 light curves.

\section{Tables and PSDs}

\begin{figure*}
\includegraphics[width=17cm]{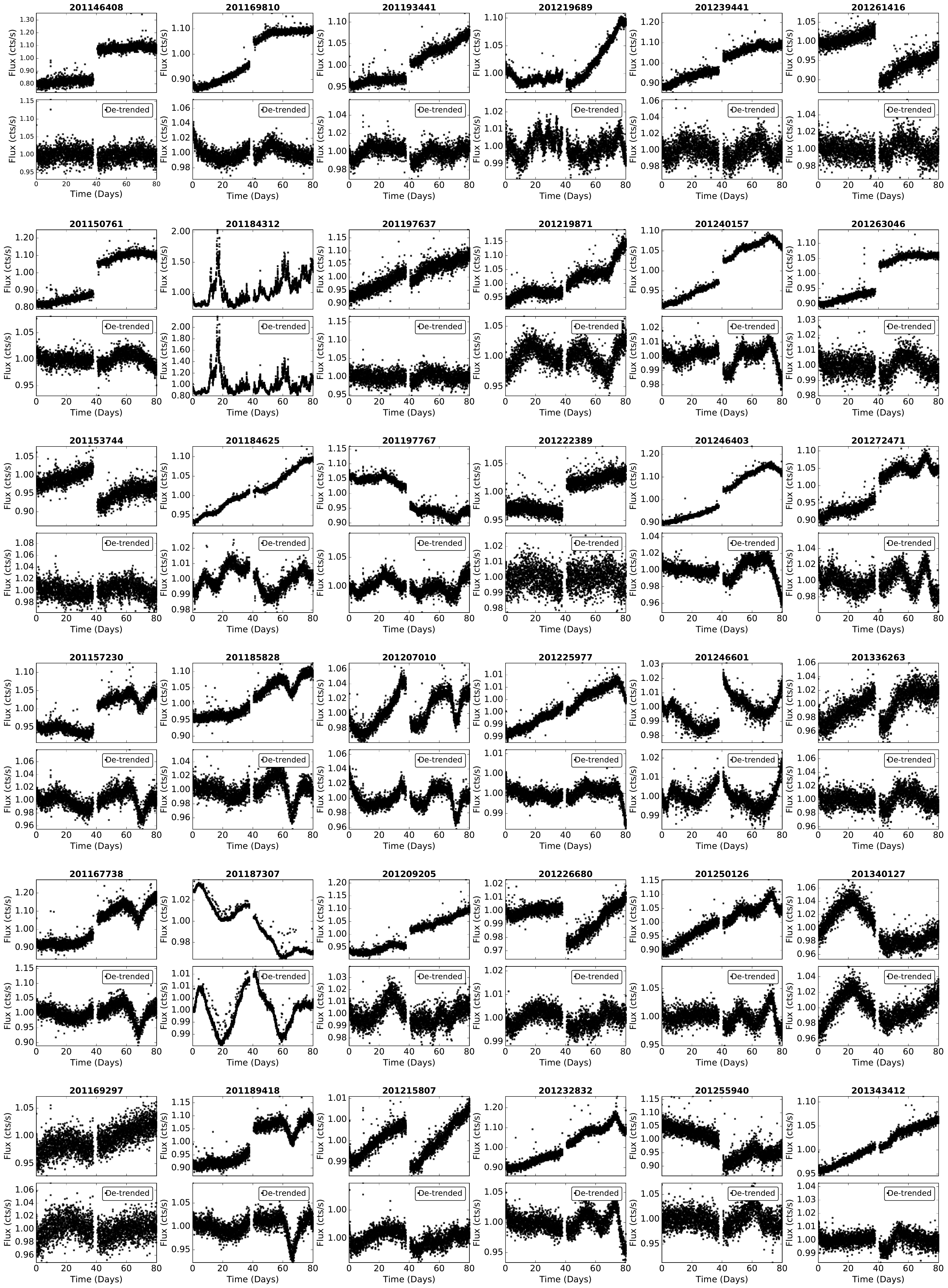}
\centering
\caption{`Optimal' and `optimal' de-trended light curves for the first 36 light curves of the sample.}
\label{Fig:panel1lcr}
\end{figure*}

\begin{figure*}
\includegraphics[width=18cm]{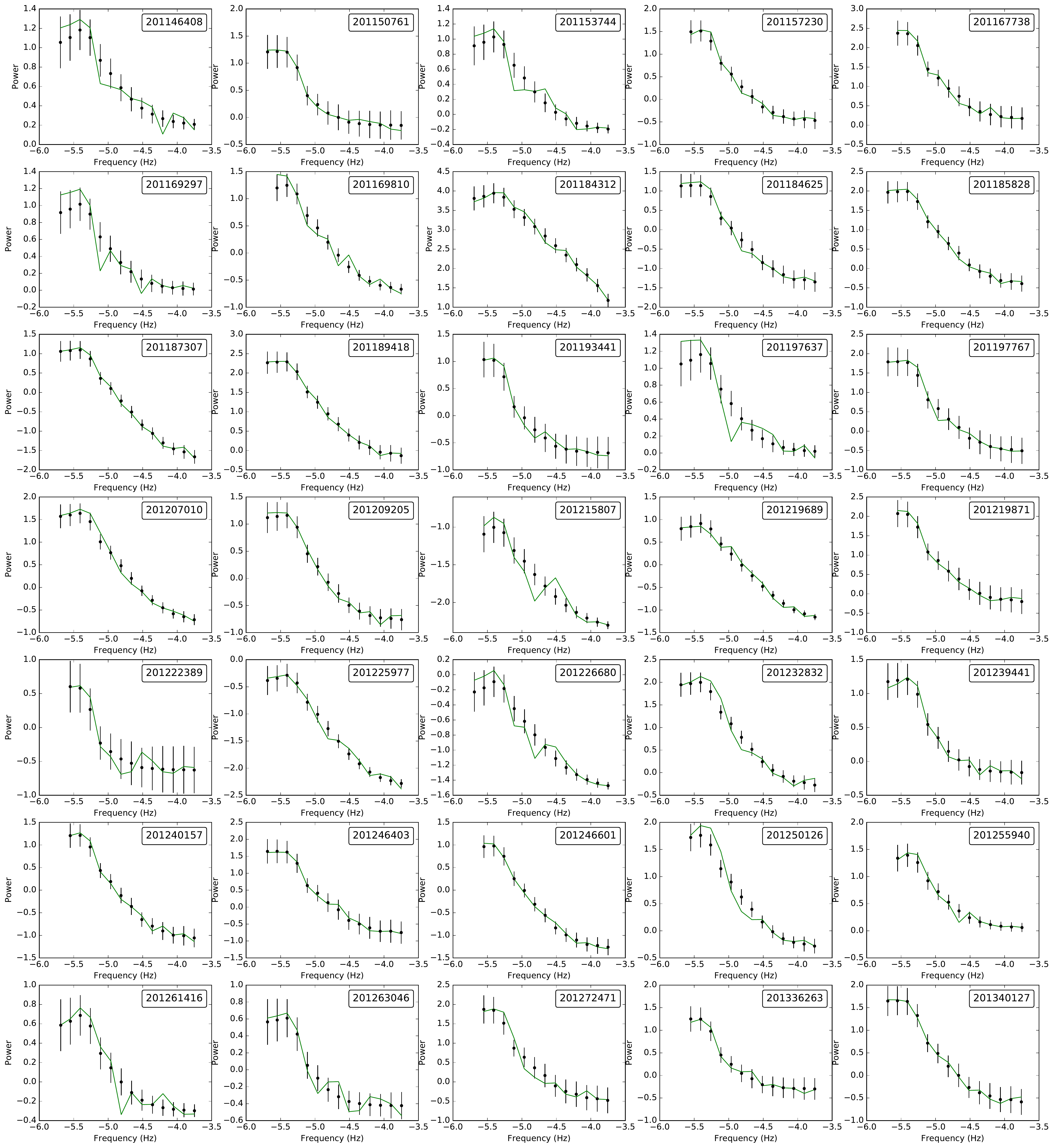}
\centering
\caption{Power spectral densities of the AGN from the MQ catalogue showing the different PSD shapes. The green solid line indicates the observed power spectra and the black filled circles with error bars are the simulated model with the highest acceptance probability.}
\label{Fig:panel1}
\end{figure*}

We include in this appendix a short version of the catalogue in Table \ref{Table:big} including 10 AGN from the MQ catalogue analysed in this work with their {\it Kepler} identifier, coordinates in J2000, physical parameters derived from the MQ catalogue and their variability properties. We have also added the estimated bolometric luminosities for AGN with known redshift up to $z\sim 2.2$, since we only have K-corrections up to that redshift. The full catalogue can be found in the on-line material. The star symbol next to the EPIC name indicates that the data must be taken with caution as the output channels are affected by the Moir\'e effect. Moreover, we include some of the light curves in Fig. \ref{Fig:panel1lcr} and their PSDs in Fig. \ref{Fig:panel1} and the rest of them can also be found on-line. We show the `optimal' light curves and below the same de-trended for comparison to asses the DVA effect. The figures of the PSDs are in logarithmic scale and are not corrected from time-dilation. 
\clearpage
\onecolumn
\begin{landscape}
\input{shorttable_sub14nov.tex}

\end{landscape}

\end{document}

%% file: latextable_14novnon.tex
205993418&335.2285&-15.0134&$2.32^{+0.79}_{-0.80}$&$1.53^{+0.82}_{-0.82}$&$1.66^{+0.57}_{-0.52}$&61.2&0.65&0.70&$0.87\pm0.08$\\
206062517&329.7345&-13.0884&$1.79^{+0.66}_{-0.66}$&$1.26^{+0.90}_{-0.90}$&$1.53^{+0.61}_{-0.61}$&90.7&1.05&1.07&$1.27\pm0.11$\\
206072629&329.8968&-12.8165&$2.71^{+0.57}_{-1.11}$&$2.05^{+1.07}_{-1.06}$&$1.92^{+0.72}_{-0.65}$&81.1&0.84&0.82&$1.11\pm0.18$\\
206454152&335.6067&-5.7358&$2.45^{+0.61}_{-0.68}$&$1.39^{+0.91}_{-0.91}$&$1.39^{+0.56}_{-0.56}$&83.6&1.82&1.84&$2.09\pm0.12$\\

%% file: shorttable_sub14nov.tex

\begin{table*}
\scriptsize
\caption{The table includes the parameters for all the MQ AGN analysed in this work: the EPIC number, coordinates in J2000, aparent magnitude in $r'$ and $b'$ band, type fo AGN  and redshift (extracted from the MQ catalogue), the negative power-law index of the simulated PSD and the error for the `optimal' aperture, the PRF aperture and the `optimal' de-trended light curves. The rest of the values correspond to the analysis of the de-trended light curves: the best normalisation and Poisson noise level obtained in the minimisation, the acceptance probability for the model, the fractional rms measured in the observed and the simulated power spectra, followed by the rest-frame fractional rms with the associated error. The last column contains the estimated bolometric luminosities in units of $10^{45}\,\rm erg\,s^{-1}$.}
\begin{tabular}{ccccccccccccccccc}
\hline\hline
EPIC & RA (deg) & DEC (deg) & Type & b${'}$ & r${'}$ & z & $\beta_{\rm opt}$ & $\beta_{\rm PRF9}$ & $\beta_{\rm det}$ & Norm & Noise & Probability ($\%$) & $\sigma_{\rm obs}$ ($\%$) & $\sigma_{\rm fit}$ ($\%$) & $\sigma_{\rm rest,fit}$  ($\%$) & $\rm L_{bol}$  ($\times10^{45}\,\rm erg\,s^{-1}$) \\
\hline
201146408&174.5331&-5.3805&q&19.4&18.4&2.00&$2.18^{+0.82}_{-1.08}$&$1.66^{+0.70}_{-0.64}$&$1.39^{+1.41}_{-1.40}$&0.46&1.54&36.9&1.12&1.06&$1.32\pm0.12$&326.25\\
201150761&174.7086&-5.2778&q&18.5&18.2&1.20&$2.84^{+0.60}_{-0.67}$&$1.79^{+1.15}_{-1.15}$&$2.71^{+0.68}_{-1.40}$&4.68&0.71&90.7&0.76&0.66&$1.29\pm0.52$&190.46\\
201153744&172.1308&-5.2070&q&18.5&18.2&1.10&$1.92^{+0.97}_{-0.35}$&$1.92^{+0.70}_{-0.31}$&$1.53^{+0.70}_{-0.69}$&1.16&0.60&44.2&0.82&0.84&$1.02\pm0.09$&153.57\\
201157230&173.9066&-5.1233&q&18.3&17.7&1.00&$2.58^{+0.71}_{-0.71}$&$2.45^{+0.00}_{-0.00}$&$2.32^{+0.69}_{-0.69}$&3.68&0.32&74.2&1.13&1.07&$1.69\pm0.38$&145.13\\
201167738&173.0793&-4.8731&q&19.2&18.5&1.40&$2.84^{+0.54}_{-0.44}$&$2.58^{+0.00}_{-0.00}$&$2.71^{+0.44}_{-0.43}$&5.79&1.38&27.8&2.58&2.54&$5.36\pm0.83$&157.59\\
201169297&172.6518&-4.8334&q&19.4&18.0&1.50&$1.79^{+1.41}_{-0.71}$&$2.18^{+1.19}_{-1.19}$&$1.66^{+1.44}_{-0.46}$&1.08&1.01&33.8&0.74&0.76&$1.03\pm0.13$&161.06\\
201169810&169.7321&-4.8214&R&18.3&17.8&0.00&$3.24^{+0.50}_{-0.50}$&$2.45^{+0.58}_{-1.03}$&$1.92^{+0.00}_{-0.00}$&2.73&0.20&8.9&0.92&0.91&$0.00\pm0.00$&0.00\\
201184312&173.9927&-4.4744&AR&18.7&17.0&0.27&$1.66^{+0.34}_{-0.44}$&$1.66^{+0.35}_{-0.38}$&$1.53^{+0.41}_{-0.41}$&1.72&-15.53&76.6&25.71&24.83&$26.45\pm2.25$&0.00\\
201184625&176.5165&-4.4670&AX&16.6&15.4&0.13&$3.11^{+0.77}_{-0.77}$&$2.97^{+0.79}_{-0.79}$&$2.58^{+0.27}_{-0.28}$&4.92&0.04&16.2&0.70&0.65&$0.71\pm0.14$&5.22\\
201185828&173.2443&-4.4373&q&19.1&18.4&1.60&$2.58^{+0.56}_{-0.70}$&$2.18^{+0.48}_{-0.37}$&$2.45^{+0.57}_{-0.57}$&4.39&0.35&79.6&1.80&1.78&$3.55\pm0.56$&5.94\\
\hline
\end{tabular}
\label{Table:big}
\end{table*}